\documentclass[aps, onecolumn, superscriptaddress]{revtex4-2}

\usepackage{amsmath,amssymb}
\usepackage{bm}
\usepackage{graphicx}
\usepackage{xcolor}

\begin{document}
\title{Pairwise scattering and bound states of spherical microorganisms}

\author{C.~Darveniza}
\affiliation{School of Mathematics and Statistics, The University of Melbourne, Parkville, Victoria 3010, Australia.}
\author{T.~Ishikawa}
\affiliation{Department of Finemechanics, Tohoku University, 6-6-01, Aoba, Aramaki, Aoba-ku, Sendai 980-8579, Japan.}
\author{T.~J.~Pedley}
\affiliation{Department of Applied Mathematics and Theoretical Physics, University of Cambridge, Centre for Mathematical Sciences, Wilberforce Road, Cambridge CB3 0WA, UK.}
\author{D.~R.~Brumley}
\email{d.brumley@unimelb.edu.au}
\affiliation{School of Mathematics and Statistics, The University of Melbourne, Parkville, Victoria 3010, Australia.}

\date{\today}

\begin{abstract}
The dynamic interactions between pairs of swimming microorganisms underpin the collective behaviour of larger suspensions, but accurately calculating pairwise collisions has typically required the use of numerical simulations in which hydrodynamic interactions are fully resolved. In this paper, we utilise analytical expressions for forces and torques acting on two closely separated spherical squirmers -- accurate to second order in the ratio of cell-cell spacing to squirmer radius -- in order to calculate their scattering dynamics. Attention is limited to squirmers whose orientation vectors lie in the same plane. We characterise the outgoing angles of pairs of bottom-heavy squirmers in terms of their incoming angles, the squirmer parameter $\beta$, and the strength of the external gravitational field, discovering transient scattering, stationary bound states, pairwise swimming motion, and circular orbits. These results compare well with full numerical solutions obtained using boundary element methods, highlighting the utility of lubrication theory. We expect these results will be useful for the foundations of mesoscale continuum models for suspensions of spherical microorganisms.
\end{abstract}

\maketitle

\section{Introduction}

Suspensions of active matter feature prominently in biological and engineered systems, and have generated great interest among fluid dynamicists and soft matter physicists \cite{Shaebani2020}. Notable examples include suspensions of swimming microorganisms \cite{Pedley1992}, collective motion of active filaments \cite{Sanchez2012} and a range of chemically- \cite{Howse:2007hc}, magnetically- \cite{Schuerle2019, Meng2021} and acoustically-driven \cite{Zhou2017} particles. Underpinning striking experimental observations such as bacterial turbulence \cite{Dombrowski2004}, stable bound states \cite{Drescher:2009vn}, and clustering of particles \cite{Thutupalli2018}, is an interplay between hydrodynamic and steric interactions, and  responses to external forces and torques \cite{Driscoll2017, Delmotte2019}.  Theoretical works of active suspensions have typically centred around continuum models -- either in the dilute limit \cite{Pedley1990} or through minimal models \cite{Wensink:2012} -- and numerical simulations which resolve interactions across multiple length scales \cite{Ishikawa2006, Ishikawa2008}. Recent work has also shown that summation of pairwise lubrication interactions between closely-separated spherical squirmers is sufficient to capture the bulk rheological properties of concentrated suspensions, across a range of packing fractions and external field strengths \cite{ishikawa2021, BrumleyPRF2019}. \\

In the pursuit of an understanding of active matter collective dynamics, experimental works have systematically examined the fluid flows generated by propulsive appendages \cite{Wei2021, Brumley2014, Brumley2015}, swimming microorganisms \cite{Drescher:2010kx, Guasto:2010ly, Drescher:2011bh, Cortese2021}, the interactions between pairs of cells \cite{Drescher:2009vn} and active droplets \cite{Lippera2021}, and the role of nearby boundaries \cite{Dehkharghani2019, Lauga2006, Hoeger2021}. Colonies of {\it Volvox} exhibit striking bound states \cite{Ishikawa2020} mediated by gravity and a bounding wall, while cells of {\it Chlamydomonas} reflect in non-trivial ways from no-slip surfaces \cite{Contino2015}. Detailed numerical studies have carefully calculated the scattering angles between pairs of squirmers in quiescent Stokes flow \cite{Ishikawa2006, Gotze2010, Llopis:2010}, and examined the role of inertia \cite{Li2016}, density stratification \cite{More2021}, and unsteady swimming \cite{Giacche2010}.  For axisymmetric interactions, it is possible to find closed form analytical solutions for the interaction between squirmers \cite{Papavassiliou2017}. However, for general interactions,  works involving closely-separated squirmers typically involve computationally-intensive simulations to calculate the fluid flows and resolve the organisms' trajectories \cite{Ishikawa2006}. The success of lubrication theory in predicting selected hydrodynamic bound states \cite{Drescher:2009vn} as well as bulk suspension rheology \cite{ishikawa2021} offers a tantalising prospect that it may be used more generally to calculate pairwise dynamics of hydrodynamically coupled microorganisms. This paper utilises analytical expressions for forces and torques between closely-separated spherical squirmers whose orientations are coplanar, calculating the hydrodynamic interactions between them and resolving their full two-dimensional trajectories. This enables rapid calculation of non-trivial scattering dynamics between pairs of arbitrarily positioned squirmers, across a range of squirming parameters (pushers vs. pullers) and for varying degrees of bottom-heaviness.

\section{Model outline}

The principal goal of this paper is to calculate the scattering dynamics of pairs of colliding spherical microorganisms of equal radius $a$. The celebrated squirmer model \cite{Blake:Squirmer} is used to capture the motility of the cells, which move due to radial and tangential velocity boundary conditions,
\begin{equation}
u_r \big|_{r=a} = \sum_{n} A_n(t) P_n (\cos \theta), \qquad u_{\theta} \big|_{r=a} = \sin \theta \sum_{n} B_n(t) W_n (\cos \theta) \label{squirmer_BC},
\end{equation}
respectively, where $\theta$ is the angle measured from the anterior of the squirmer, $P_n$ is the $n^{\text{th}}$ Legendre polynomial, and $W_n$ is defined as
\begin{equation}
W_n(\cos \theta) = \frac{2}{n(n+1)} P_n' (\cos \theta).
\end{equation}

The swimming speed of a solitary squirmer is related to the primary squirming mode, $V_s = 2B_1/3$, and $\beta = B_2 / B_1$ captures the stresslet strength. We consider squirmers with zero radial velocity ($A_n(t) = 0 \; \forall \; n$), and although the analysis can be completed to arbitrary order in $B_n$, we will truncate the tangential squirming modes to $n=1,2$. The overall velocity boundary condition is therefore given by 
\begin{equation}
u_{\theta} = \tfrac{3}{2} V_s \sin \theta (1 + \beta \cos \theta). \label{squirmer_u_theta}
\end{equation}
The cartesian coordinates of the spheres' centres $\bm{x}_i$ and orientations $\bm{p}_i$ (for $i=1,2$) are considered to lie in the $x$-$y$ plane. The spheres may be bottom heavy, so that when the swimming direction, $\bm{p}_i$, is not vertical, the squirmer experiences a gravitational torque, 
\begin{equation}
\bm{T}_i^{grav} = - \rho v h  \bm{p}_i \times \bm{g}, \label{T_grav}
\end{equation}
where $v$ and $\rho$ are the cell volume and average density, respectively; $h$ is the displacement of the centre of mass from the geometric centre; and $\bm{g}$ is the gravitational vector with $|\bm{g}|=g$. At each instant in time, the squirmers each experience forces and torques due to their intrinsic swimming motion (propulsion); hydrodynamic drag due to the background fluid; mutual hydrodynamic interactions; and the action of gravity. We also include a repulsive interparticle force to prevent the spheres from overlapping,
\begin{equation}
\bm{F}^{rep} = \eta_0 a^2 \kappa_1 \kappa_2 \frac{ \exp (- \kappa_2 \epsilon)}{ 1 -  \exp (- \kappa_2 \epsilon)} \hat{\bm{r}}, \label{repulsive_force}
\end{equation}
where $\hat{\bm{r}}$ is the unit vector along the line of centres and $\epsilon$ is the minimum dimensionless clearance between the two cells, $\epsilon a = | \bm{x}_1 - \bm{x}_2 | - 2a$. We apply $\kappa_1 = 10$ and $\kappa_2 = 100$.\\

To calculate the hydrodynamic interactions between two squirmers, we utilise linearity of the Stokes equations and solve for the interaction between a squirmer and a no-slip sphere. Figure~\ref{fig_schematic}(a) shows the arrangement of the squirmer-sphere pair used to derive the forces and torques due to squirming in the lubrication limit. The solution in this frame can be mapped to any relative position/orientation of squirmers through a linear transformation. Without loss of generality, we apply the squirming-sphere boundary condition on sphere 1 and zero velocity boundary condition on sphere 2. The forces and torques acting on the two spheres were previously calculated explicitly \cite{BrumleyPRF2019}, and are given by

\begin{align}
\begin{split}
F_x^{(1)} &= -\frac{4}{5} \mu \pi a \ \bm{p} \cdot \bm{e}_x \frac{\lambda (\lambda+4)}{(\lambda+1)^2} \sum_n B_n W_n \big( -\bm{p} \cdot \bm{e}_z \big)  \big( \log \epsilon + \mathcal{O}(1) \big), \\
F_z^{(1)} &= -9 \mu \pi a \frac{\lambda^2}{(\lambda+1)^2} \sum_n \bigg[ B_n W_n \big( - \bm{p} \cdot \bm{e}_z \big) \bm{p} \cdot \bm{e}_z + \frac{1}{2} B_n W_n' \big( - \bm{p} \cdot \bm{e}_z \big) ( \bm{p} \cdot \bm{e}_{x} )^2 \bigg] \big( \log \epsilon + \mathcal{O}(1) \big), \\
T_y^{(1)} &= \frac{16 \lambda}{5 (\lambda+1)} \mu \pi a^2 \ \bm{p} \cdot \bm{e}_x \sum_n B_n W_n \big( -\bm{p} \cdot \bm{e}_z \big) \big( \log \epsilon + \mathcal{O}(1) \big), \\
T_y^{(2)} &= \frac{4 \lambda^2}{5 (\lambda+1)} \mu \pi a^2 \ \bm{p} \cdot \bm{e}_x \sum_n B_n W_n \big( -\bm{p} \cdot \bm{e}_z \big) \big( \log \epsilon + \mathcal{O}(1) \big).
\label{forces_and_torques}
\end{split}
\end{align}
The above expressions are valid for all tangential squirming modes, but henceforth we will consider only $n=1,2$. We also consider spheres of equal radii, $\lambda=1$.

Consistent with previous works \cite{BrumleyPRF2019, Pedley2016}, the total dimensional force $\bm{F}_i$ and torque $\bm{T}_i$ on each squirmer are scaled according to $\bar{\bm{F}}_i = \bm{F}_i / (\eta_0 \pi a)$ and $\bar{\bm{T}}_i = \bm{T}_i /(\eta_0 \pi a^2)$, so that they each have units of velocity. We introduce a dimensionless number $G_{bh}$, defined as
\begin{equation}
G_{bh} = \frac{4 \pi \rho g a h}{3 \eta_0 V_s},
\end{equation} 
which measures the relative importance of gravitational torques compared to viscous torques for the squirmers. The gravitational torque on each squirmer can then be written as
\begin{equation}
\bar{\bm{T}}_i^{grav} = - \frac{1}{\pi} V_s G_{bh} \bm{p}_i \times \hat{\bm{g}}.
\end{equation}
Each squirmer is subject to a propulsive force of constant magnitude, parallel to its orientation vector, according to
\begin{equation}
\bar{\bm{F}}_i^{prop} = 6 V_s \bm{p}_i, \label{F_prop}
\end{equation}
which ensures propulsion of an isolated squirmer, at the prescribed swimming speed $V_s$. The complete resistance formulation for the two spheres in Stokes flow is then given by
\begin{equation}
\bm{R} \cdot \left( \begin{array}{c}
\bm{V} \\
a \bm{\omega} \\
\end{array} \right) = - \left( \begin{array}{c}
\bar{\bm{F}}^{sq} + \bar{\bm{F}}^{rep} + \bar{\bm{F}}^{prop} \\
\bar{\bm{T}}^{sq} + \bar{\bm{T}}^{grav} \\
\end{array} \right), \label{resistance_formulation}
\end{equation}
where the resistance matrix is $\bm{R} = \bm{R}^{sq-sq} + \bm{R}^{drag}$. The matrix $\bm{R}^{sq-sq}$ incorporates the hydrodynamic forces and torques acting on both spheres, arising from their linear and angular velocities. These expressions correspond to squeezing, shearing, and rotation of no-slip spheres in the lubrication limit, and can be found in full in \cite{BrumleyPRF2019} and \cite{KimAndKarrila:Microhydrodynamics}. The diagonal matrix $\bm{R}^{drag}$ represents the hydrodynamic drag due to the translation and rotation of an isolated sphere in Stokes flow, and is given by 
\begin{equation}
\bm{R}^{drag} = \left( \begin{array}{c|c}
-6 \bm{I}_{4} & \bm{0} \\
\hline
\bm{0} & -8 \bm{I}_{2} \\
\end{array} \right),
\end{equation}
where $\bm{I}_{n}$ is the $n \times n$ identity matrix. On the right hand side of Eq.~\eqref{resistance_formulation}, the forces, $\bar{\bm{F}}^{sq}$, and torques, $\bar{\bm{T}}^{sq}$, arise due to the squirming motion of each sphere. We refer the reader to \cite{BrumleyPRF2019} for detailed expressions noting that a change of reference frame must be made for each pair, in order for the expressions to apply.
The resistance formulation in Eq.~\eqref{resistance_formulation} can be written in the form
\begin{equation}
\bm{R}(\mathbf{x}) \cdot \dot{\mathbf{x}} + \bm{S}(\mathbf{x}) = \bm{0},
\label{resistance_formulation2}
\end{equation}
where $\mathbf{x}$ is a $6 \times 1$ vector containing the position and orientation of both spheres (confined to lie in the $x$-$y$ plane), and both $\bm{R}$ and $\bm{S}$ depend on $\mathbf{x}$. This enables us to calculate the linear and angular velocities of both squirmers based on their respective positions and orientations. However, we note that interparticle interactions -- mediated through $\bm{R}^{sq-sq}$, $\bar{\bm{F}}^{sq}$, $\bar{\bm{T}}^{sq}$ and $\bar{\bm{F}}^{rep}$ --  occur only in the lubrication limit. The squirming forces and torques exhibit a logarithmic dependence \cite{BrumleyPRF2019} on the interparticle spacing, $\epsilon = (| \bm{x}_1 - \bm{x}_2 | - 2a)/a$, so $\epsilon = 1$ is a natural choice for the limit for lubrication interactions.

For widely-separated squirmers ($\epsilon > 1$), the resistance matrix is diagonal, and therefore Eq.~\eqref{resistance_formulation} can be solved explicitly to yield
\begin{equation}
\bm{V}_i = V_s \bm{p}_i, \qquad 
a \bm{\omega}_i = - \frac{1}{8 \pi} V_s G_{bh} (\bm{p}_i \times \hat{\bm{g}}).
\end{equation}
As expected, solitary squirmers will swim at a speed $V_s$ in a direction $\bm{p}_i$ which can change due to the presence of gravitational torques. 

\newpage
\section{Results for non-bottom heavy squirmers} \label{section_nonBH}

\begin{figure}[htp]
\begin{center}
	\includegraphics [width=135mm]{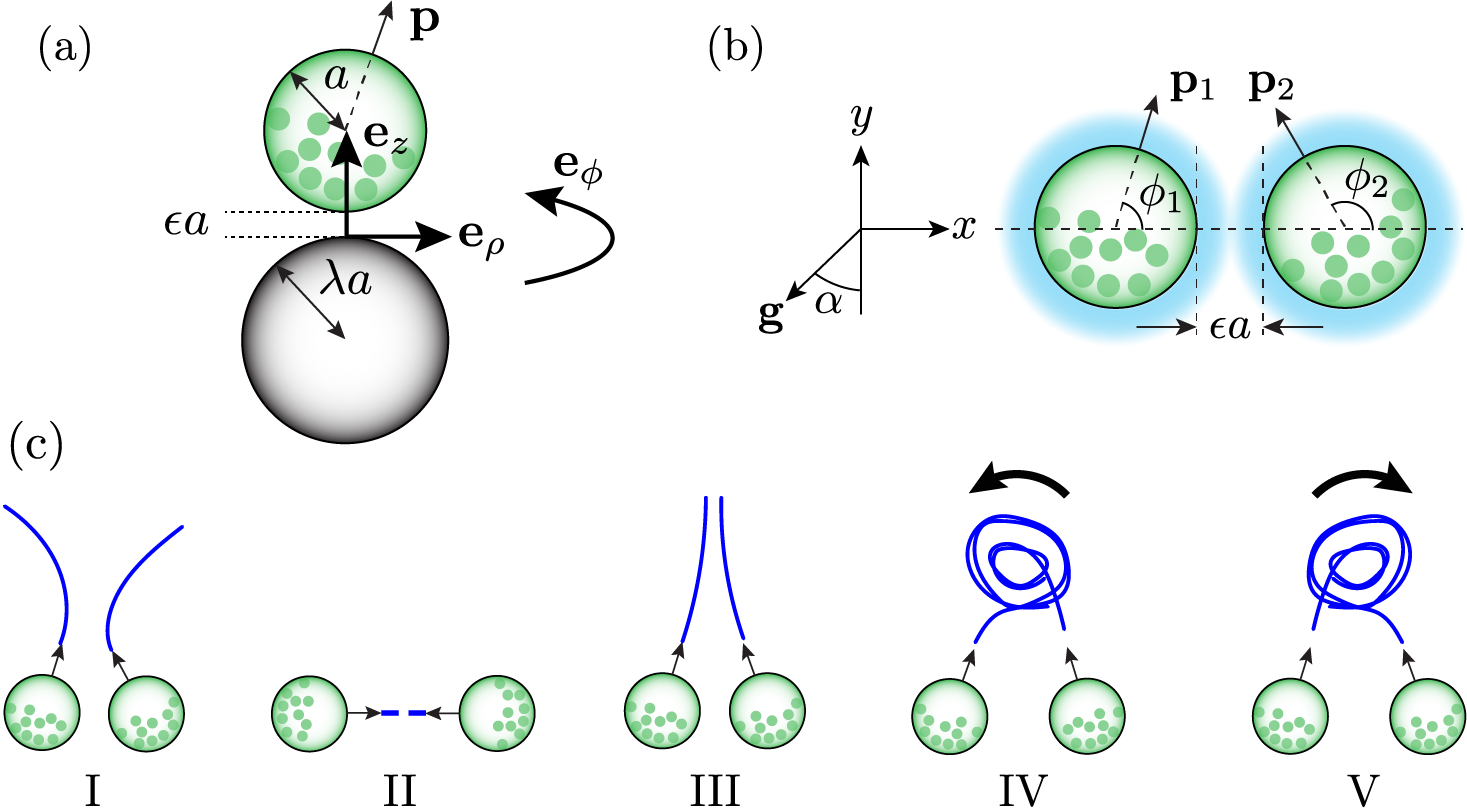} 
\caption{ Calculating squirmer scattering dynamics. (a) The hydrodynamic interactions between a steady spherical squirmer (radius $a$, swimming direction $\bm{p}$) and a no-slip sphere (radius $\lambda a$) are calculated in the lubrication limit $\epsilon \ll 1$, resulting in explicit expressions for the hydrodynamic forces and torques, Eq.~\eqref{forces_and_torques}. The origin of the coordinate system is located on the surface of sphere 2 closest to sphere 1. The vector $\bm{e}_{\rho}$ points radially in the $x$-$y$ plane, and the vector $\bm{e}_{\phi}$ is the azimuthal direction. 
(b) For two squirmers with initial orientations $\phi_1$ and $\phi_2$, respectively, separated by distance $(| \bm{x}_1 - \bm{x}_2 | - 2a)/a = 1$, the trajectories are calculated using Eq.~\eqref{resistance_formulation}. (c) The resultant dynamics can be categorised into (I) non-zero scattering over a finite time; (II) stationary standoff; (III) pairwise swimming; (IV) anticlockwise orbiting; (V) clockwise orbiting. } \label{fig_schematic}
\end{center}
\end{figure}

In the absence of gravitational torques, widely-separated squirmers swim with a constant velocity, $V_s \bm{p}_i$. Deviations from these trajectories only occur if the spacing is sufficiently close ($\epsilon \leq 1$) so as to facilitate hydrodynamic interactions. In order to characterise the pairwise scattering dynamics of non-bottom heavy squirmers, it is therefore sufficient to consider pairs whose initial  non-dimensionalized separation is exactly $\epsilon = 1$.  The cases where squirmers begin with a larger separation will either have the squirmers fail to come into close contact with one another and continue in straight lines, or reach a point where $\epsilon =1$ so that the remaining path is described by our results. We consider the initial positions of the spheres to lie on the $x$-axis, as shown in Fig.~\ref{fig_schematic}(a), with orientations $\phi_i$ measured anticlockwise from the positive $x$-axis. Equation~\ref{resistance_formulation2} is integrated directly over $t \in [ 0, 200 ]$, to find the position and orientation of both squirmers as functions of time. The duration of these simulations is long enough to ensure that squirmers had either departed from each other’s vicinity ($\epsilon >1$) and thereafter drifted away from one another in straight lines, or entered a state that they would stay in indefinitely, such as a steady state, closed orbiting pattern or periodic motion. At the end of each simulation, the final angles $\Phi_i$ were recorded for each squirmer $i$. We also recorded the duration of the interaction, $\tau$, defined as the length of time for which the two squirmers interact hydrodynamically ($\epsilon \leq 1$). Unless otherwise stated, all positions and distances will be considered as being non-dimensionalized with respect to the sphere radius $a$. 

Although it might appear natural to express the outgoing squirmer angles $\Phi_i$ in terms of incoming angles $\phi_i$, we note that half of the possible initial conditions result in squirmers immediately swimming apart from one another (e.g., for $\text{mod} (\phi_2, 2\pi)  < \text{mod} (\phi_1, 2\pi) $). Furthermore, these configurations do not correspond to initial collisions between straight-swimming squirmers, since reversing the direction of time would result in hydrodynamic interactions. When characterising the scattering properties of the squirmers, it is therefore most appropriate to consider the outgoing angles $\Phi_i$ in terms of the difference, $\Delta = (\phi_2-\phi_1)/\pi$, and sum, $\Sigma = (\phi_2+\phi_1)/\pi$, of the incoming angles. Figure~\ref{fig_schematic}(c) classifies the different dynamics observed in the numerical simulations.

\subsection{Neutral squirmers} \label{results_neutral}

\begin{figure}[htp]
\begin{center}
	\includegraphics [width=125mm]{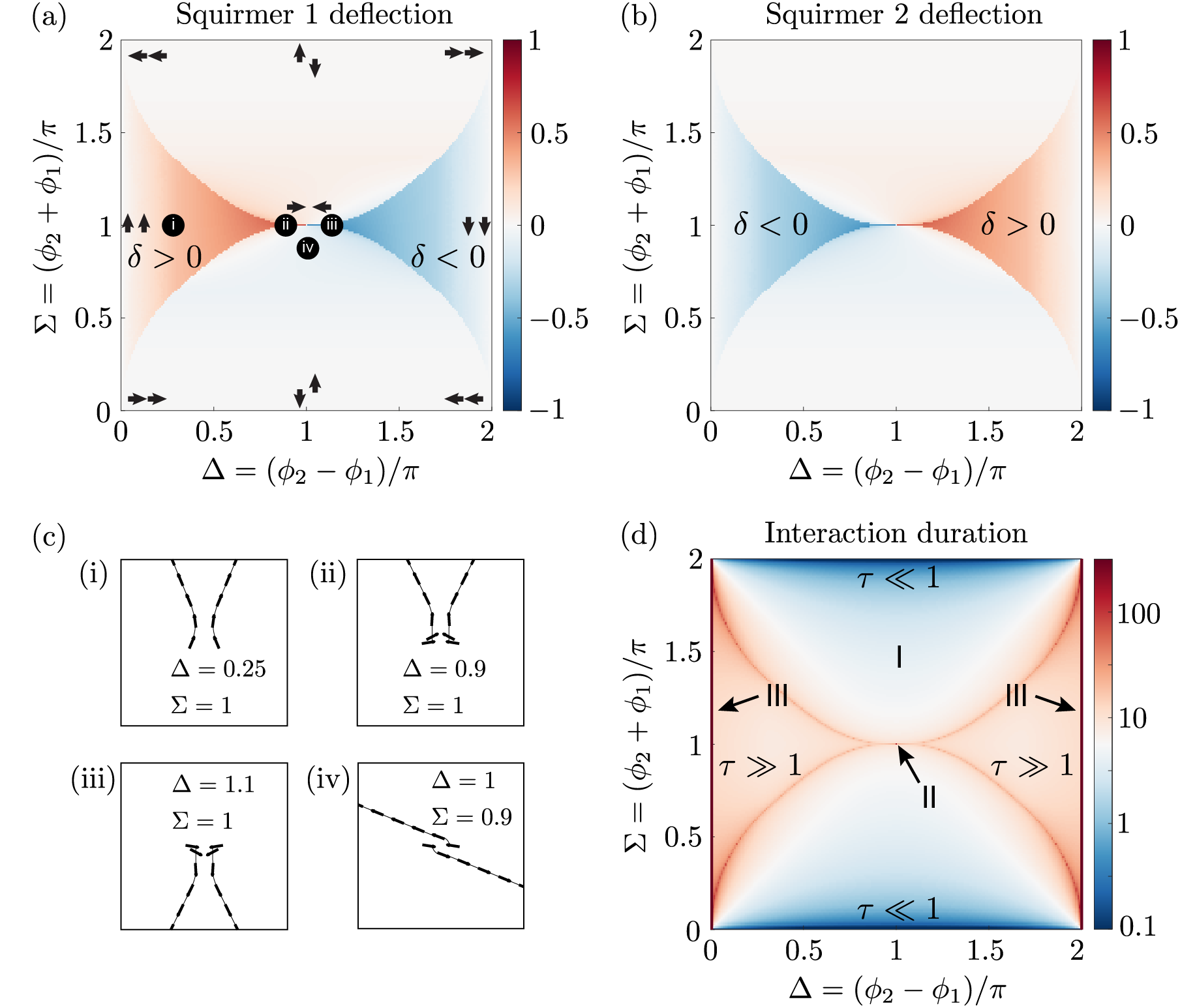} 
\caption{Hydrodynamic interactions between neutral squirmers ($\beta=0$). Deflection angle $\delta_i = (\Phi_i - \phi_i)/\pi$ for (a) squirmer 1 and (b) squirmer 2, as a function of initial configuration $\Delta = (\phi_2-\phi_1)/\pi$ and $\Sigma = (\phi_2+\phi_1)/\pi$. (c) Example trajectories, in which symbols (i-iv) correspond to labels in a. (d) The interaction duration, $\tau$, represents the time taken for the squirmers to separate ($\epsilon>1$). For $\Delta=0$, $\Delta=2$ or $\Delta=\Sigma=1$ the squirmers exhibit a permanent bound state ($\tau = \infty$). All other configurations result in scattering.} \label{fig_beta0_scattering}
\end{center}
\end{figure}

We begin by examining the simplest case of two interacting neutral squirmers ($\beta=0$). For any initial conditions $(\phi_1,\phi_2)$ -- or equivalently $(\Delta, \Sigma)$ -- it is sufficient to consider the outgoing angle of squirmer 1, namely $\Phi_1$. The deflection angle of squirmer 2 can be found by symmetry, since the squirmers are situated in an infinite domain. However, for illustrative purposes, both angles are shown alongside one another in this section. Figures~\ref{fig_beta0_scattering}(a) and (b) show the deflection angles $\delta_1 = (\Phi_1-\phi_1)/\pi$ and $\delta_2 = (\Phi_2-\phi_2)/\pi$, respectively, in terms of the initial conditions $\Delta$ and $\Sigma$. Pairs of black arrows in (a) depict the example initial squirmer orientations.

There are several configurations which give rise to permanent bound states. We refer the reader to Fig.~\ref{fig_schematic}(c) for classification of the states. For $\Delta =0$ (or equivalently $\Delta=2$), the two neutral squirmers swim parallel alongside one another (Case III). When both $\Delta=1$ and $\Sigma=1$, the two squirmers are initially pointed directly at one another, and as time progresses, they swim together to form a stationary standoff (Case II). All of these bound states are depicted in Fig.~\ref{fig_beta0_scattering}(d) with extended interaction duration, $\tau = \infty$. While the value of exactly $\Delta=1$ and $\Sigma=1$ results in a permanent standoff, this configuration is highly unstable. The greatest deflections are observed close to this point, when $\Sigma=1$ and $0 < | \Delta - 1 | \ll 1$. Under these conditions, the extrapolation of the squirmer's trajectories from $t=0$ would cross one another. The squirmers undergo substantial deflections beyond $\pi/2$ ($\delta \approx \pm 0.60$), and both swim off in either the positive $y$-direction (Fig.~\ref{fig_beta0_scattering}(c)ii, Movie 1) or negative $z$-direction (Fig.~\ref{fig_beta0_scattering}(c)iii), depending on the sign of $\Delta - 1$. 

For squirmers whose initial orientations would not lead to intersecting trajectories, the squirmers slide past one another, and depart in opposite directions with only a minor deflection from the initial orientations (e.g., Fig.~\ref{fig_beta0_scattering}(c)iv). Although the interaction duration of cases (iii) and (iv) are not substantially different (11.4\ s and 5.0\ s respectively), the former case exhibits 8 times greater deflection ($\delta_1= -0.60$ vs. $\delta_1=-0.074$), the key determinant being whether the initial configuration of squirmers would see their future trajectories intersecting or not.

Despite the weak deflection for $0<\Delta \ll 1$, the two squirmers swim with almost collinear trajectories, and therefore remain close to one another for an extended period of time (see Fig.~\ref{fig_beta0_scattering}(d)). Along the top and bottom boundaries of Fig.~\ref{fig_beta0_scattering}(d) ($\Sigma \approx 0$ or $\Sigma \approx 2$ with intermediate $0<\Delta<2$), the interaction duration is extremely short-lived because squirmers tend to swim off quickly in distinct directions.

\subsection{Varying $\beta$} \label{results_vary_beta}

When $\beta \neq 0$, the two squirmers exhibit fore aft asymmetry in their tangential velocity boundary condition, with the magnitude of $\beta$ capturing the effective stresslet strength. We now examine the deflection, $\delta_1$, of squirmer 1 as a function of the initial difference, $\Delta$, and sum, $\Sigma$, of the squirmer orientations. For brevity, we omit the deflection of squirmer 2, noting that it can be obtained by symmetry from the results for $\delta_1$.

\begin{figure}[htp]
\begin{center}
	\includegraphics [width=120mm]{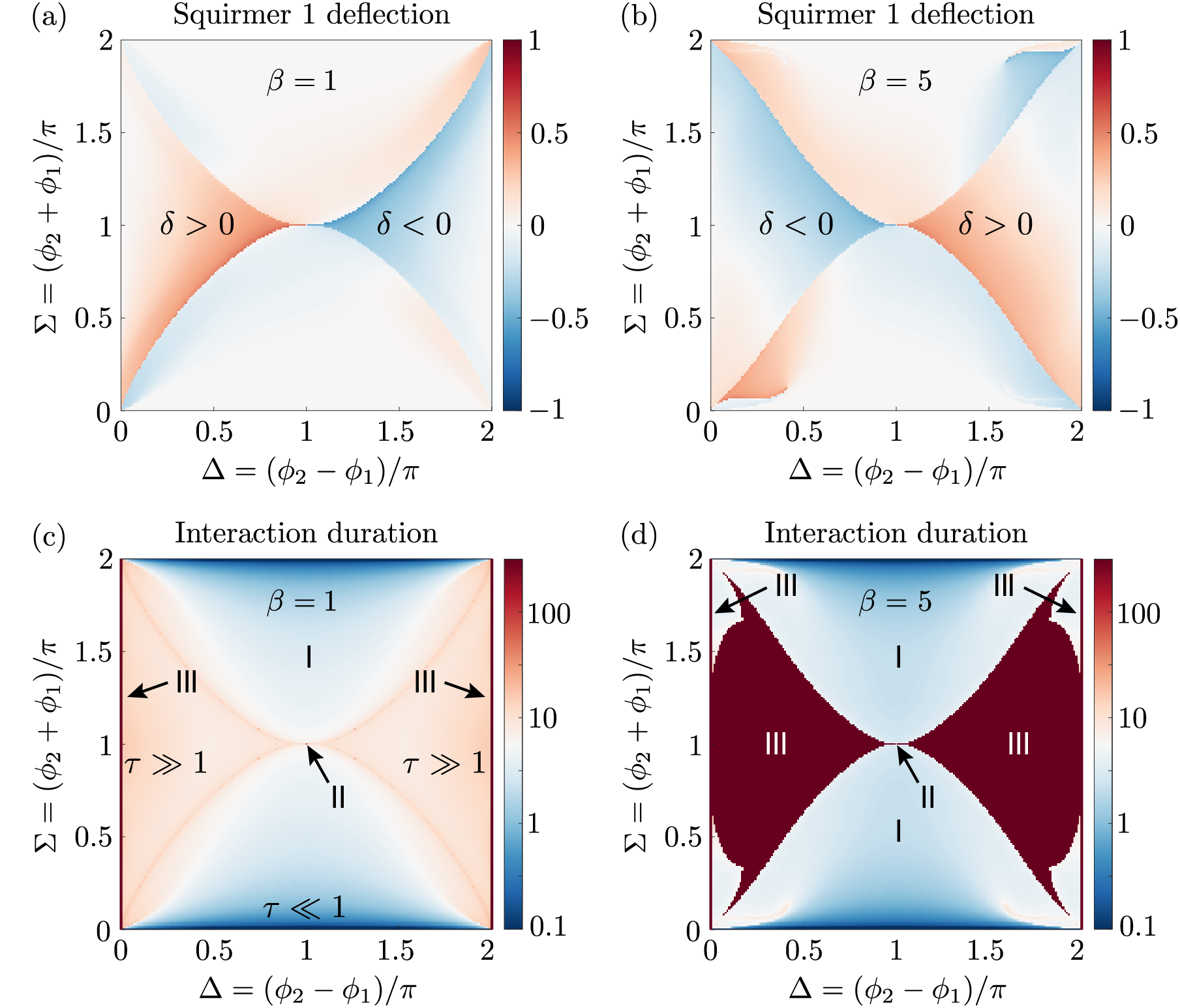} 
\caption{Hydrodynamic interactions between pullers ($\beta>0$). Deflection angle for squirmer 1, $\delta_1 = (\Phi_1 - \phi_1)/\pi$, for (a) $\beta=1$ and (b) $\beta = 5$. Results are shown as a function of initial configuration $\Delta = (\phi_2-\phi_1)/\pi$ and $\Sigma = (\phi_2+\phi_1)/\pi$. The interaction duration, $\tau$, shown for (c) $\beta=1$ and (d) $\beta = 5$ represents the time taken for the squirmers to separate ($\epsilon>1$). Solid red regions correspond to permanent trapping. The classification of dynamics (I, II, etc.) follows the definitions in Fig.~\ref{fig_schematic}(c).} 
\label{fig_puller_scattering}
\end{center}
\end{figure}

To begin with, we consider $\beta>0$, corresponding to pairs of pullers. Figures~\ref{fig_puller_scattering}(a) and (c) show the deflection of squirmer 1 for $\beta=1$ and the associated duration of the hydrodynamic interaction, respectively. The duration of the interaction is extremely similar to the neutral squirmer case (see Fig.~\ref{fig_beta0_scattering}(d)), with permanent steady states only for $\Delta=0$, $\Delta=2$ (parallel swimming) or $\Delta=\Sigma=1$ (standoff). All other configurations result in scattering. As in the $\beta=0$ case, squirmer 1 tends to rotate away from squirmer 2 (i.e. $\delta_1 > 0$ when $\Delta \lesssim 1$ and $\delta_1 < 0$ when $\Delta \gtrsim 1$). The strongest deflections for $\beta=1$ are accompanied by relatively long interaction times (see Fig.~\ref{fig_puller_scattering}(c)). 

Although the scattering angles for strong pullers, $\beta=5$ (see Fig.~\ref{fig_puller_scattering}(b)) are similar to the $\beta=1$ case, the dynamics of the squirmers are qualitatively different.  Approximately 36\,\% of initial incoming configurations ($0 \leq \Delta \leq 2$ and $0 \leq \Sigma \leq 2$) result in permanent bound states. Figure~\ref{fig_puller_scattering}(d) reveals values of $\Delta$ and $\Sigma$ for which this permanent trapping occurs. In the red shaded region (denoted III), the squirmers each adopt a constant final orientation, with the pair swimming together in a linear fashion.  As before, the single symmetric case $\Delta=\Sigma=1$ results in a stationary standoff (case II), and $\Delta=0, 2$ also results in parallel swimming.

\begin{figure}[htp]
\begin{center}
	\includegraphics [width=130mm]{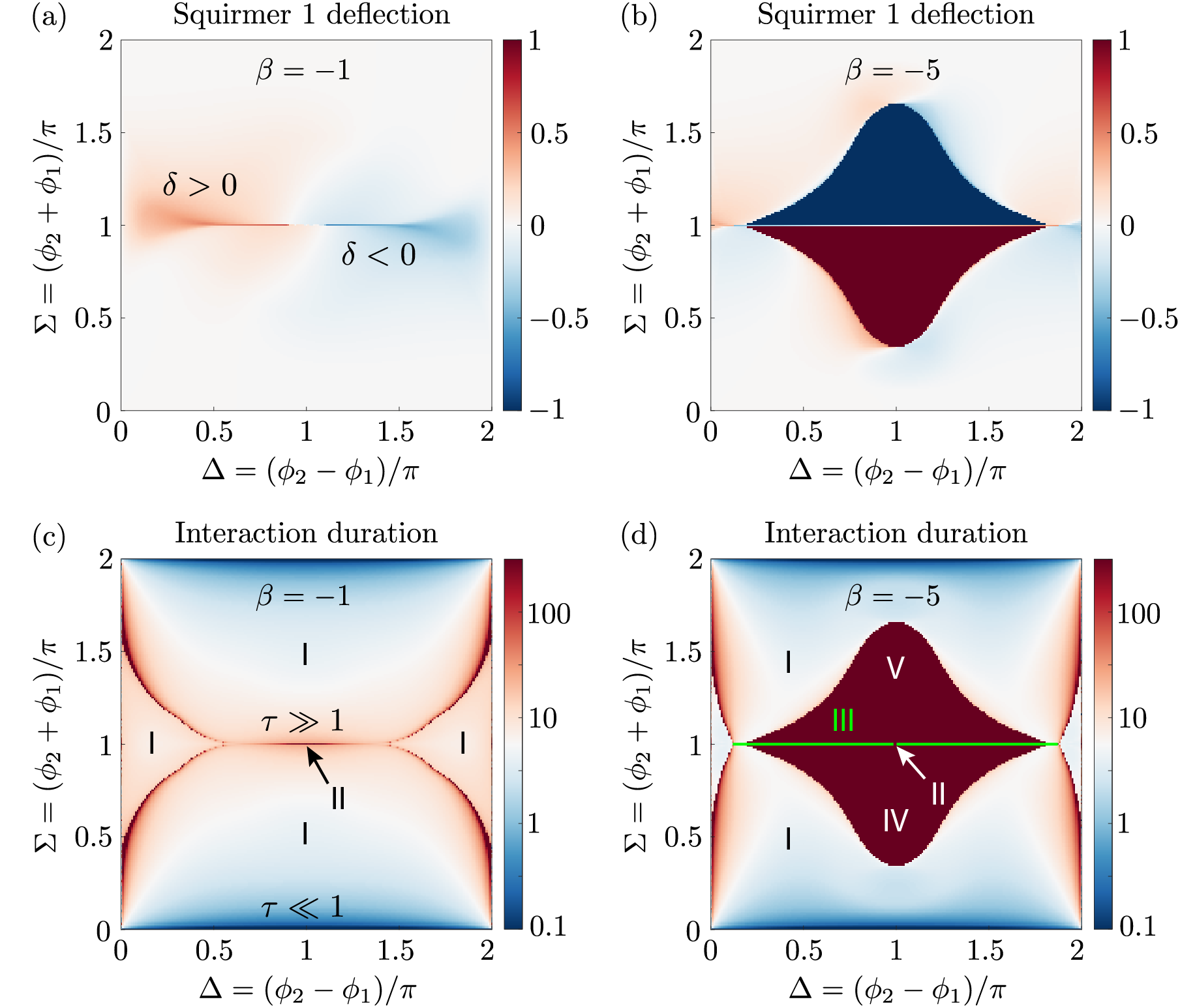} 
\caption{Hydrodynamic interactions between pushers ($\beta<0$). Deflection angle for squirmer 1, $\delta_1 = (\Phi_1 - \phi_1)/\pi$, for (a) $\beta=-1$ and (b) $\beta = -5$. Results are shown as a function of initial configuration $\Delta = (\phi_2-\phi_1)/\pi$ and $\Sigma = (\phi_2+\phi_1)/\pi$. The interaction duration, $\tau$, shown for (c) $\beta=-1$ and (d) $\beta = -5$ represents the time taken for the squirmers to separate ($\epsilon>1$).} 
\label{fig_pusher_scattering}
\end{center}
\end{figure}

We now consider the results corresponding to pushers ($\beta<0$). As in all previous cases, two squirmers directly facing one another ($\Delta=\Sigma=1$) result in a permanent standoff. However, the broader results are markedly different from the neutral and puller cases. The deflection angles for $\beta=-1$ and $\beta=-5$ are shown in Figs.~\ref{fig_pusher_scattering}(a) and (b) respectively. Within region I of these figures, squirmers again exhibit a finite interaction time, resulting in deflected squirmers with independent outgoing angles. However, in comparison with neutral squirmers and pullers, the deflection angle does not vary as dramatically across the range of initial conditions. For $\beta=-1$, strong scattering occurs only for $\Sigma \approx1$  (close to symmetrically oriented squirmers).  This value corresponds to squirmers whose anterior hemispheres collide. The tangential slip velocity in this anterior region tends to rotate the two squirmers toward one another, resulting in strong deflections. The dark red regions of Fig.~\ref{fig_pusher_scattering}(c) represent prolonged, but not permanent, trapping. With the exception of the singular standoff and parallel swimming cases, squirmers with $\beta=-1$ ultimately depart.

For strong pushers, $\beta=-5$, we observe qualitatively different dynamics. For 23\,\% of the possible initial conditions, the squirmers exhibit permanent bound states (red shaded region of Fig.~\ref{fig_pusher_scattering}(d)). When $\Sigma=1$, the squirmer configuration is symmetric about the $y$-axis, and so the bound state corresponds to pairwise swimming in a straight line (case III). However, for $|\Sigma-1| > 0$, the squirmers do not approach symmetrically, and instead may adopt a bound state in which cells orbit one another in an anticlockwise (case IV) or clockwise (case V) manner, depending on the sign of $\Sigma-1$.  These orbiting states are characterised by squirmers whose orientations differ by exactly $\pi$, but whose swimming axes are slightly offset. We note that deflection angles in Fig.~\ref{fig_pusher_scattering}(b) for these permanently bound squirmers do not represent ``final'' equilibrium outgoing angles, but the instantaneous deflection of the circular orbit at the end of the simulation.

\begin{figure}[htp]
\begin{center}
	\includegraphics [width=160mm]{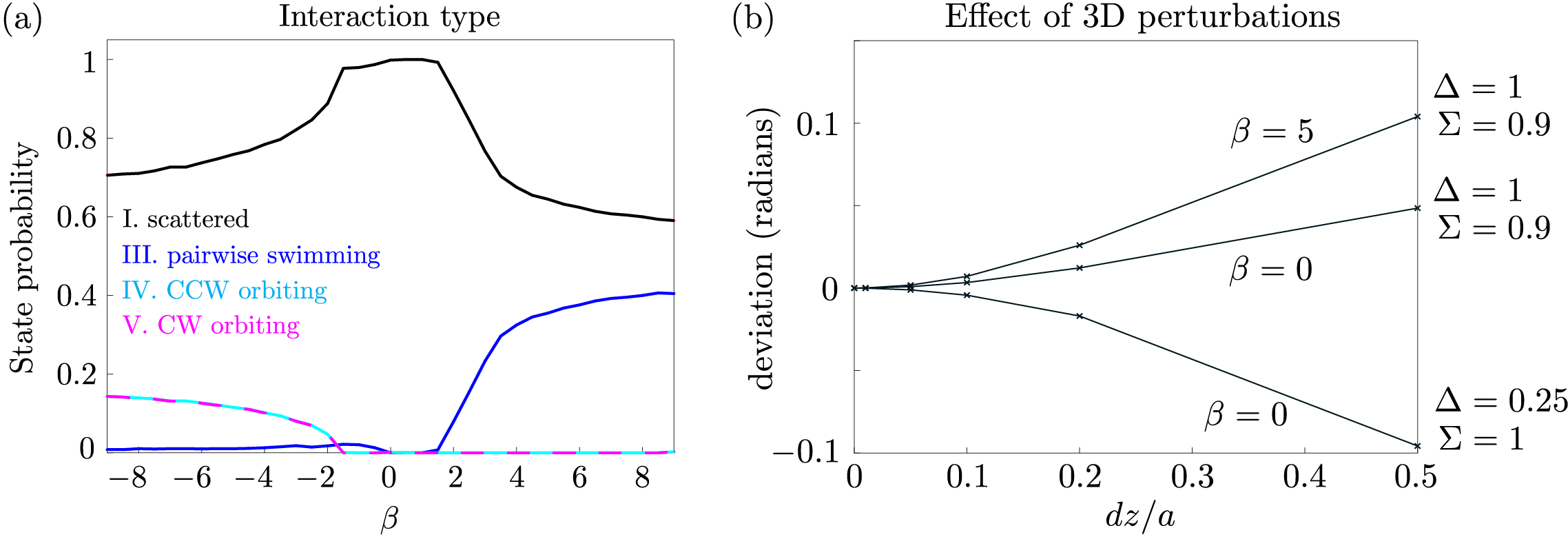} 
\caption{ (a) Characterisation of the squirmer interactions as a function of $\beta$. For each value of $\beta$, the probability is calculated that squirmers will exhibit transient scattering, indefinite side-by-side swimming, counter-clockwise orbiting, or clockwise orbiting. These results are shown for $G_{bh}=0$. (b) Changes in the scattering angles due to a 3D offset in the squirmers initial positions (normalised by sphere radius). Results are shown for several representative collisions and squirmer parameters. } 
\label{fig_vary_beta}
\end{center}
\end{figure}

The results in Figs.~\ref{fig_beta0_scattering}-\ref{fig_pusher_scattering} reveal qualitatively different interactions across the discrete values of $\beta$ studied so far ($\beta=0, \pm 1, \pm 5$). Aside from specific symmetries, neutral squirmers did not exhibit hydrodynamic trapping. Conversely, pullers and pushers tended to exhibit pairwise linear swimming (case III) and periodic orbiting (cases IV-V) respectively. To further investigate the dependence of dynamics on stresslet strength, we calculated -- for various values of $\beta$ -- the probability that a squirmer pair will exhibit scattering, pairwise swimming or periodic orbiting, from the initial conditions $\Delta, \Sigma \in ( 0, 2)$. The results are presented in Fig.~\ref{fig_vary_beta}. The dynamics for specific values of $\Delta=0,2$ (parallel swimming) and $\Delta=\Sigma=1$ (squirmer standoff) are present across all values of $\beta$ studied. Moreover, they constitute infinitesimally small regions of $\Delta \times \Sigma$ space, and are therefore overlooked in the probability calculation. This is accomplished by using a uniform mesh for $\eta \leq \Delta \leq 2-\eta$ and $\eta \leq \Sigma \leq 2-\eta$ with an even number of points and $0< \eta \ll 1$.

The results of Fig.~\ref{fig_vary_beta}(a) generalise the results found earlier in Figs.~\ref{fig_beta0_scattering}-\ref{fig_pusher_scattering}. Neutral squirmers do not exhibit hydrodynamic bound states (with the exception of specific $\Delta=0,2$, $\Delta=\Sigma=1$ cases), but rather scatter with finite interaction time. The likelihood of trapping increases monotonically with $|\beta|$. For pullers ($\beta>0$), these states consist exclusively of pairwise linear swimming, while pushers ($\beta<0$) tend to exhibit orbiting motion, with no net drift in the orbit centre. By symmetry, clockwise and counterclockwise orbiting are equally likely.

 To examine the sensitivity of the results to out-of-plane perturbations, we calculated the 3D trajectories of squirmer pairs whose initial orientations were confined to lie in the $x$-$y$ plane, but positions were offset by distance $dz$ in the $z$-direction. The boundary element method \citep{Ishikawa2006} was utilised to calculate trajectories for various initial orientations, squirmer parameters and values of $dz$. In each case, we compared the outgoing scattered angles for $dz>0$ with the equivalent values in the case $dz=0$. In most cases, the scattering angles differed only by up to $\sim 6^{\circ}$, even with $z$-offset up to half the squirmer radius (see Fig.~\ref{fig_vary_beta}(b)). However, we note that bound states, e.g. Fig.~\ref{fig_schematic}(c)III, can be disrupted by small perturbations in the $z$-direction, since they require highly symmetric configurations to persist.

\vspace{-15pt}
\subsection{Comparison with full numerical model}

\begin{figure}[htp]
\begin{center}
	\includegraphics [width=170mm]{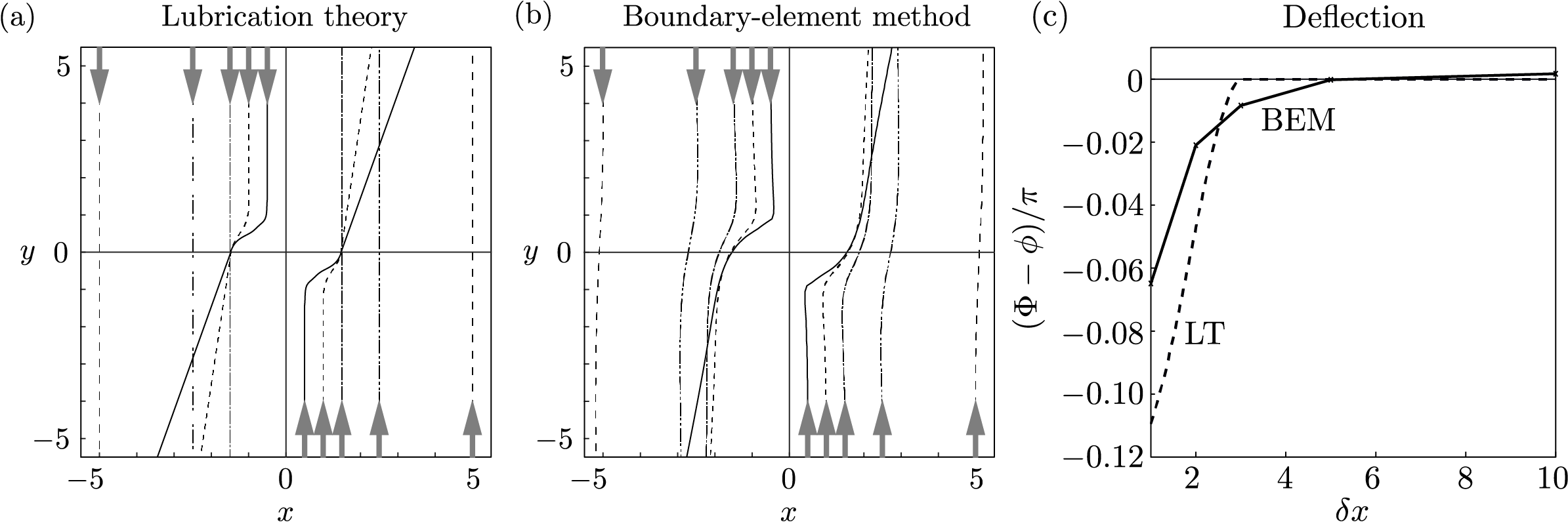} 
\caption{ Trajectories of squirmers subject to (a) lubrication-only hydrodynamic interactions (see Eq.~\eqref{resistance_formulation2}) and (b) full hydrodynamic interactions. Parameters are given by $\beta=5$, $G_{bh}=0$. (c) Deflection angle experienced by each squirmer, as a function of initial spacing, $\delta x$. } 
\label{fig_numerical_comparison}
\end{center}
\end{figure}

To this point, the pairwise dynamics of the squirmers have been calculated with hydrodynamic interactions limited to those arising from lubrication regions ($\epsilon \leq 1$). Here we compare calculated trajectories of squirmers with those arising from full boundary element method (BEM) simulations of Ishikawa {\it et al.}  \citep{Ishikawa2006} but with the short range repulsive force switched on (see. Eq.~\eqref{repulsive_force}). We first examine collisions between squirmers whose initial velocity vectors are opposite in direction. The initial coordinates of the squirmers are chosen to match Ref~\cite{Ishikawa2006}, with $\bm{x}_1 = (- \delta x/2,5)$, $\bm{x}_2 = (+\delta x/2,-5)$, $\phi_1 = -\pi/2$ and $\phi_2 = +\pi/2$. This separation is sufficiently large so as to ensure that initial interactions in the fully coupled case are negligible, and squirmers in both cases will swim in straight lines. Figure~\ref{fig_numerical_comparison} illustrates the trajectories of squirmers approaching one another with $\delta x = 1, 2, 3, 5, 10$, for lubrication theory (panel a) and boundary element method (panel b). In the absence of gravitational torques, squirmers subject only to lubrication theory interactions (LT) swim in straight lines until $\epsilon = (| \bm{x}_1 - \bm{x}_2 | - 2a)/a \leq 1$. It follows that for $\delta x>3$, LT squirmers do not experience any near-field interactions, and thus do not deviate from their linear trajectories (see Fig.~\ref{fig_numerical_comparison}(a)). Conversely, the BEM squirmers for $\delta x>3$ do exhibit deviations from their initial trajectories due to longer range interactions. However, despite the small horizontal translation, we note that the swimming direction in these cases remains essentially unchanged. For initial conditions with $\delta x<3$, we observe good agreement between the LT and BEM methods. In both cases, the magnitude of the angular deflection increases as $\delta x$ is reduced. 

The results of Fig.~\ref{fig_numerical_comparison}(a) can  be interpreted through Fig.~\ref{fig_puller_scattering}(b), by considering the point at which the two squirmers first come within a distance $\epsilon \leq 1$ of one another (valid only for $\delta x \leq 3$). This is equivalent to the case where $\Delta=(\phi_2-\phi_1)/\pi = 1$ and $\Sigma = (\phi_2+\phi_1)/\pi = 1 - 2\arcsin(\delta x /3)/\pi$. Figure~\ref{fig_puller_scattering}(d) confirms that all such cases of off-axis approach result in finite interactions and scattering of squirmers, except the symmetric head-on case of $\delta x=0$ (corresponding to $\Sigma=1$), which results in a stand-off. 
 Figure~\ref{fig_numerical_comparison}(c) shows the angular deflection experienced by each squirmer, as a function of the initial spacing, $\delta x$. Results are shown for both the lubrication theory (LT) and boundary element method (BEM), and exhibit good agreement across different spacings. 

\begin{figure}[htp]
\begin{center}
	\includegraphics [width=125mm]{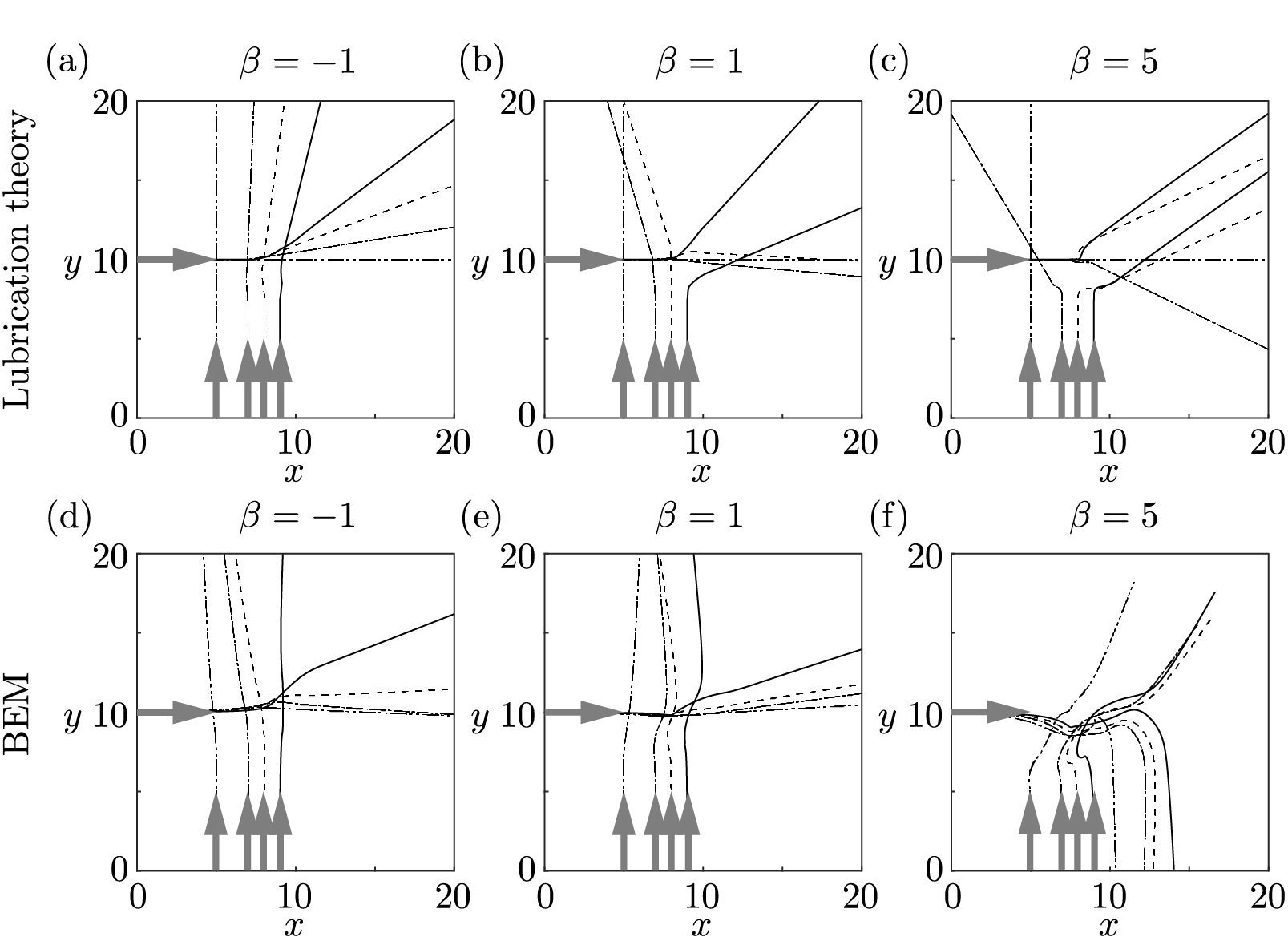} 
\caption{Trajectories of squirmers subject to (a-c) lubrication-only hydrodynamic interactions and (d-f) full hydrodynamic interactions. The initial conditions are given by $\bm{x}_1 = (0, 10)$, $\bm{x}_2 = (10-\delta x,0)$, $\phi_1 = 0$ and $\phi_2 = \pi/2$ where $\delta x = 1$, 2, 3, 5. Results are plotted for $\beta=-1$, 1, 5, with  $G_{bh}=0$ in all cases.} 
\label{fig_numerical_comparison2}
\end{center}
\end{figure}

We further examined the scattering dynamics of squirmer pairs whose initial velocity vectors differ by $\pi/2$ (see Fig.~\ref{fig_numerical_comparison2}). There is good agreement between all trajectories for $\beta=-1$. For $\beta=1$ (Fig.~\ref{fig_numerical_comparison2}(b,e)), the agreement is reasonable except for the case where $\delta x=1$ (solid lines): the LT simulations predict squirmers whose paths do not cross, while the BEM model exhibits crossing trajectories. The differences are more pronounced for the $\beta=5$ case (Fig.~\ref{fig_numerical_comparison2}(c,f)), and the reasons for this are discussed later.

\newpage
\section{Results for bottom heavy squirmers}

Many microorganisms, including {\it Volvox} \cite{Drescher:2009vn}, {\it Chlamydomonas} \cite{Kessler:1985a} and {\it Heterosigma akashiwo} \cite{Sengupta2017}, exhibit a displacement between their geometric centre and centre of mass, resulting in an effective bottom-heaviness. We now consider the effect of an external gravitational torque, as outlined in Eq.~\eqref{T_grav}, on the pairwise dynamics of spherical squirmers. To exhibit representative dynamics, we focus on the case where $\alpha=0$, so that the direction of gravity is perpendicular to the line joining the spheres' centres (see Fig.~\ref{fig_schematic}), though in principle this could be varied. An isolated squirmer whose initial orientation $\bm{p}$ is not parallel to the gravitational field $\bm{g}$, will experience an external torque which balances the viscous torques of rotation, to gradually reorient the squirmer until $\bm{p}$ is antiparallel to $\bm{g}$. However, for pairs of squirmers, the hydrodynamic interactions provide additional forces and torques which can facilitate richer dynamics.

\begin{figure}[htp]
\begin{center}
	\includegraphics [width=\textwidth]{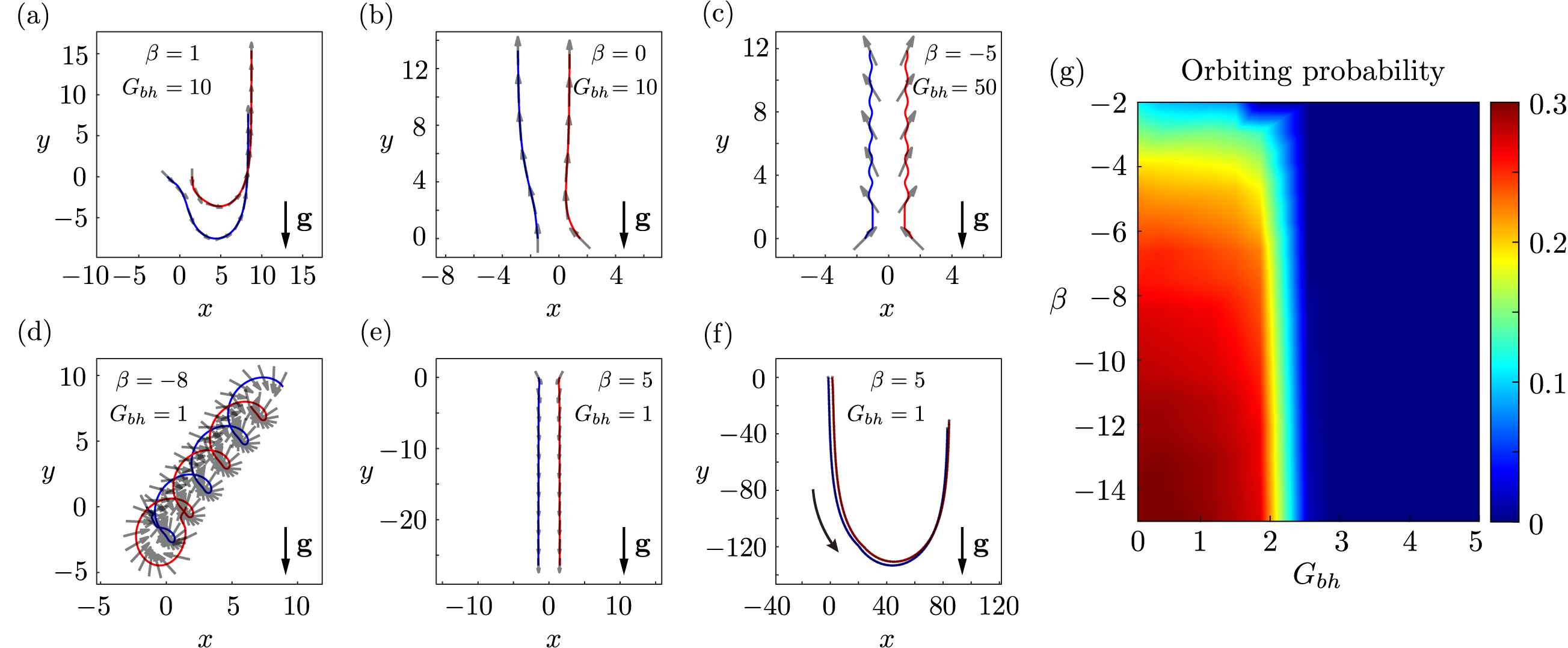} 
\caption{ Calculated trajectories for bottom-heavy squirmers, starting from initial conditions $\bm{x}_1 = (-1.5, 0)$ and $\bm{x}_2 = (+1.5, 0)$. (a-f) The trajectories of squirmers 1 and 2 are shown in blue and red, respectively. The orientation of each squirmer is shown at intervals of $\Delta t = 4$, using translucent black arrows. (g) For each value of $\beta$ and $G_{bh}$, the probability of the bottom-heavy squirmer pair being in a permanent drifting orbit state (see panel d) is shown. } \label{fig_bottom_heavy}
\end{center}
\end{figure}

For sufficiently strong $G_{bh}$, the gravitational torque is expected to dominate the dynamics, resulting in upward swimming motion of the squirmer pair. Figures~\ref{fig_bottom_heavy}(a-b) show representative results for $G_{bh}=10$, with $\beta=1$ and $\beta=0$ respectively. The hydrodynamic interactions temporarily compete with gravitational torques, but ultimately the squirmers' swimming direction is righted against gravity. Figure~\ref{fig_bottom_heavy}(c) reveals an interesting bound state in which both stronger gravity ($G_{bh}=50$) and greater stresslet ($\beta=-5$) result in fluctuations in the spacing between the squirmers. With the squirmers splayed away from one another, they interact solely through their posterior regions. The pusher boundary condition on each squirmer provides a torque which opposes that due to bottom-heaviness. Based on their orientations, the cells have a tendency to swim away from one another. However, this is balanced by the fact that pushers ($\beta<0$) draw fluid away from their equator, and therefore attract one another in this region.  

For smaller values of $G_{bh}$, an interplay between the hydrodynamic interactions and gravitational torques can give rise to non-trivial bound states. For strong pushers, pairs of squirmers may exhibit orbiting motion, with the pair moving diagonally against gravity.  Figure~\ref{fig_bottom_heavy}(d) shows an example of one such trajectory, which we refer to as a ``drifting orbit''.  The orbiting direction depends on the initial angles of the two squirmers. These results can be reconciled with Fig.~\ref{fig_pusher_scattering}(b,d), in which strong non-bottom heavy pushers may orbit one another. If gravity is sufficiently weak, it may bias the mean direction of the orbiting pair without disrupting the perpetual bound state.

To investigate the parameter combinations giving rise to drifting orbits, we calculated the squirmer trajectories over a range of $G_{bh}$, $\beta$, $\Delta$ and $\Sigma$ values. For given $G_{bh}$ and $\beta$, we calculated the proportion of initial orientations that  resulted in drifting orbits, the results of which are summarised in Fig.~\ref{fig_bottom_heavy}(g). In Section~\ref{results_vary_beta}, we saw that for $\beta=-5$, $G_{bh}=0$, approximately 23\% of incoming angles result in permanent trapping (red shaded region of Fig.~\ref{fig_pusher_scattering}(d)), effectively all of which were orbiting dynamics. By increasing $G_{bh}$ from zero, we see that the overall probability of orbiting decreases, up to a value of $G_{bh} \approx 2.5$, where orbiting no longer occurs, and the squirmers separate. \\

Highly specific symmetries can give rise to hydrodynamic bound states of downward swimming squirmers. The squirmers in Fig.~\ref{fig_bottom_heavy}(e), oriented symmetrically about the $y$-axis with $\Delta=1.7$ and $\Sigma = 1$, are subject to gravitational torques which would tend to rotate the squirmers towards one another. However, this is balanced by the viscous torques generated by pullers interacting through anterior contact regions. Importantly, the results are qualitatively different to those in Fig.~\ref{fig_bottom_heavy}(c), since pullers draw fluid from their poles to the equator, and therefore `repel' one another when aligned close to parallel. This perfectly balances the propulsive force, giving rise to an equilibrium state. As is the case for an isolated squirmer, this downward swimming state is unstable, with a small asymmetry in the initial swimming angles ultimately resulting in a U-turn (see Fig.~\ref{fig_bottom_heavy}(f) for $\Delta=1.75$, $\Sigma=1.05$). However, we note that the bottom-heavy squirmer pair is capable of swimming downwards for a considerable amount of time before gravitational torques eventually reorient the pair. Specifically, the bound pair in Fig.~\ref{fig_bottom_heavy}(f) swim 4.5 times further downward than they would as isolated squirmers, revealing that hydrodynamic trapping can provide `inertia' from the influence of external torques.

\section{Discussion}

The main focus of this paper has been to calculate the dynamics of colliding spherical squirmers, over a range of incoming angles, squirming parameters and external gravitational fields. Recent work has demonstrated that the bulk rheology of a concentrated suspension of spherical squirmers can be quantitatively captured by considering only lubrication interactions \cite{ishikawa2021} -- that is, hydrodynamic interactions only between cells and their nearest neighbours. Here we have explicitly calculated the swimming trajectories of pairs of squirmers subject to these short-range hydrodynamic interactions, revealing good agreement with simulations in which hydrodynamic interactions are fully resolved using boundary element methods \cite{Ishikawa2006}. Our results reveal a rich range of dynamics, including transient scattering, stalling motion, pairwise linear and circular orbits. 

We have examined the pairwise dynamics for squirmers with initial angles $\phi_1$ and $\phi_2$, measured anticlockwise from the positive $x$-axis. However, it is more appropriate to consider the difference, $\Delta = (\phi_2-\phi_1)/\pi$, and sum, $\Sigma = (\phi_2+\phi_1)/\pi$, of the incoming angles over the range $ [ 0, 2] $, since this excludes initial configurations in which squirmers swim directly away from one another. The simulations were run for sufficiently long so as to determine whether cells would interact transiently, and swim apart with scattered angles, $\Phi_1$ and $\Phi_2$, or remain bound indefinitely. In all cases, we calculated the squirmers' change in orientation $\delta_i = (\Phi_i-\phi_i)/\pi$ due to the interaction. Neutral squirmers exhibit deflections of magnitude up to $\approx 2\pi/3$ ($\delta = \pm 0.64$, see Fig.~\ref{fig_beta0_scattering}). The results exhibit strong discontinuities in scattering angles across the line which determines whether squirmers slide past one another or rotate to swim as a pair. Importantly, all interactions are transient, except for highly specific symmetric cases ($\Delta=0$, $2$ and $\Delta=\Sigma=1$).

Introducing a stresslet through a non-zero value of $\beta$ results in qualitatively different results. For strong pullers ($\beta=5$), squirmers may exhibit pairwise bound states, in which cells swim together with a constant velocity. For both $\beta=1$ and $\beta=5$, Fig.~\ref{fig_puller_scattering} reveals strong discontinuities in the outgoing angle emerging with respect to incoming angles. It is also worth noting that as $\beta$ increases, the sign of the deflection switches for many of the initial configurations. Pairs of pushers exhibited the richest behaviour, with a large fraction of the initial conditions resulting in hydrodynamic bound states (see Fig.~\ref{fig_pusher_scattering}). For strong pushers ($\beta=-5$), we observed pairwise linear motion, or circular orbiting motion in a direction that depends on the squirmers' initial orientations (clockwise for $\Sigma>1$, anticlockwise for $\Sigma<1$).

Including the effect of bottom-heaviness results in external torques which must balance those from the squirming motion and the relative motion of the two spheres. Many initial configurations ultimately converge towards a pair of upward swimming cells, but other observations include periodic orbiting with drift, downward swimming, or oscillating bound states.

The good agreement between lubrication theory results (LT) and boundary element method (BEM) highlights the utility of the present method in efficiently calculating the scattering dynamics. Indeed, Ishikawa {\it et al.} showed that BEM and LT predict similar forces and torques for squirmer separations as large as $\epsilon \sim 1$ \cite{Ishikawa2006}. The ability for lubrication theory to capture hydrodynamic bound states and orbiting dynamics of {\it Volvox carteri} -- an experimental realisation of a spherical squirmer -- near no-slip boundaries has been well-documented \cite{Drescher:2009vn, Ishikawa2020}, but the scattering dynamics have not been systematically explored. Near-field hydrodynamics are also known to dominate the collective motions of ellipsoidal squirmers \cite{Kyoya2015}. Other works involving individual squirmers  swimming near a wall \cite{Lintuvuori2016, Chaithanya2021}  have demonstrated hydrodynamic oscillations. For that case, in which the ratio of minimum clearance to squirmer radius was $\sim 0.2$, lubrication theory interactions were not expected to be very accurate \cite{Lintuvuori2016}. However, in the present model of colliding squirmers, the minimum separation is significantly smaller (e.g., $\epsilon \sim 0.028$ for $\beta=0$). 
 For pairs of squirming spheres subject to axisymmetric interactions, Papavassiliou \& Alexander were able to use the reciprocal theorem to find exact solutions for the dynamics \cite{Papavassiliou2017}. However, this was only a partial solution, since the non-axisymmetric components have not been determined. Moreover, the authors did not consider the effects of external torques, for example due to gravity. The present lubrication theory scheme provides greater generality across squirmer configurations and external torques, albeit with some compromise of accuracy, and the boundary element method represents an accurate benchmark for assessing solutions. 

There has been considerable work examining the interactions of other kinds of microswimmers, e.g., phoretic particles \cite{Varma2019}, active droplets \cite{Lippera2021}, active particles \cite{Thutupalli2018}  and mixed suspensions \cite{Bardfalvy2020}.  However, in addition to hydrodynamic interactions, chemical trails can be far-reaching, and therefore affect the far-field dynamics. By examining only the near field hydrodynamic interactions, we have overlooked the structure of the flow field generated by individual squirmers. This could be important for strong pullers, which generate closed streamlines in their wake. A nearby squirmer could experience flow in the opposite direction to the tangential boundary condition. This likely contributes to the discrepancy between LT and BEM methods for $\beta=5$ in the specific case of orthogonal approach, Fig.~\ref{fig_numerical_comparison2}. However, since organisms generally approach one another and collide with anterior hemispheres, this is likely to be important only in limited cases.

Although the focus of this paper has been on the interaction between identical squirmers whose orientations are confined to the same plane, this work could be generalised to non-identical squirmers colliding with arbitrary orientation vectors. Another possible extension to this work involves considering axial rotation of squirmers while swimming, with particular application to {\it Volvox} \cite{Pedley2016}. Finally, the classification of pairwise collisions presented in this paper could be used to predict the collective motion of non-dilute suspensions of squirmers, in which interactions are dominated by pairwise collisions. This could be accomplished either using look-up-tables (e.g. results of Figs.~\ref{fig_beta0_scattering}-\ref{fig_pusher_scattering}), or through developing a collision integral formulation \cite{Nassios2016} akin to the study of rarefied gases. \\

\vspace{-10pt}
\begin{acknowledgements}
C.D. was supported by the Bernard Hoskins Scholarship, Newman College, University of Melbourne. T.I. was supported by the Japan Society for the Promotion of Science Grant-in-Aid for Scientific Research (JSPS KAKENHI Grant No. 17H00853, 17KK0080, 21H04999 and 21H05308). D.R.B. was supported by an Australian Research Council (ARC) Discovery Early Career Researcher Award DE180100911.
\end{acknowledgements}


\begin{thebibliography}{48}%
\makeatletter
\providecommand \@ifxundefined [1]{%
 \@ifx{#1\undefined}
}%
\providecommand \@ifnum [1]{%
 \ifnum #1\expandafter \@firstoftwo
 \else \expandafter \@secondoftwo
 \fi
}%
\providecommand \@ifx [1]{%
 \ifx #1\expandafter \@firstoftwo
 \else \expandafter \@secondoftwo
 \fi
}%
\providecommand \natexlab [1]{#1}%
\providecommand \enquote  [1]{``#1''}%
\providecommand \bibnamefont  [1]{#1}%
\providecommand \bibfnamefont [1]{#1}%
\providecommand \citenamefont [1]{#1}%
\providecommand \href@noop [0]{\@secondoftwo}%
\providecommand \href [0]{\begingroup \@sanitize@url \@href}%
\providecommand \@href[1]{\@@startlink{#1}\@@href}%
\providecommand \@@href[1]{\endgroup#1\@@endlink}%
\providecommand \@sanitize@url [0]{\catcode `\\12\catcode `\$12\catcode
  `\&12\catcode `\#12\catcode `\^12\catcode `\_12\catcode `\%12\relax}%
\providecommand \@@startlink[1]{}%
\providecommand \@@endlink[0]{}%
\providecommand \url  [0]{\begingroup\@sanitize@url \@url }%
\providecommand \@url [1]{\endgroup\@href {#1}{\urlprefix }}%
\providecommand \urlprefix  [0]{URL }%
\providecommand \Eprint [0]{\href }%
\providecommand \doibase [0]{https://doi.org/}%
\providecommand \selectlanguage [0]{\@gobble}%
\providecommand \bibinfo  [0]{\@secondoftwo}%
\providecommand \bibfield  [0]{\@secondoftwo}%
\providecommand \translation [1]{[#1]}%
\providecommand \BibitemOpen [0]{}%
\providecommand \bibitemStop [0]{}%
\providecommand \bibitemNoStop [0]{.\EOS\space}%
\providecommand \EOS [0]{\spacefactor3000\relax}%
\providecommand \BibitemShut  [1]{\csname bibitem#1\endcsname}%
\let\auto@bib@innerbib\@empty
\bibitem [{\citenamefont {Shaebani}\ \emph {et~al.}(2020)\citenamefont
  {Shaebani}, \citenamefont {Wysocki}, \citenamefont {Winkler}, \citenamefont
  {Gompper},\ and\ \citenamefont {Rieger}}]{Shaebani2020}%
  \BibitemOpen
  \bibfield  {author} {\bibinfo {author} {\bibfnamefont {M.~R.}\ \bibnamefont
  {Shaebani}}, \bibinfo {author} {\bibfnamefont {A.}~\bibnamefont {Wysocki}},
  \bibinfo {author} {\bibfnamefont {R.~G.}\ \bibnamefont {Winkler}}, \bibinfo
  {author} {\bibfnamefont {G.}~\bibnamefont {Gompper}},\ and\ \bibinfo {author}
  {\bibfnamefont {H.}~\bibnamefont {Rieger}},\ }\bibfield  {title} {\bibinfo
  {title} {{Computational models for active matter}},\ }\href@noop {}
  {\bibfield  {journal} {\bibinfo  {journal} {Nature Reviews Physics}\ }\textbf
  {\bibinfo {volume} {2}},\ \bibinfo {pages} {181} (\bibinfo {year}
  {2020})}\BibitemShut {NoStop}%
\bibitem [{\citenamefont {Pedley}\ and\ \citenamefont
  {Kessler}(1992)}]{Pedley1992}%
  \BibitemOpen
  \bibfield  {author} {\bibinfo {author} {\bibfnamefont {T.~J.}\ \bibnamefont
  {Pedley}}\ and\ \bibinfo {author} {\bibfnamefont {J.~O.}\ \bibnamefont
  {Kessler}},\ }\bibfield  {title} {\bibinfo {title} {{Hydrodynamic Phenomena
  in Suspensions of Swimming Microorganisms}},\ }\href@noop {} {\bibfield
  {journal} {\bibinfo  {journal} {Annual Review of Fluid Mechanics}\ }\textbf
  {\bibinfo {volume} {24}},\ \bibinfo {pages} {313} (\bibinfo {year}
  {1992})}\BibitemShut {NoStop}%
\bibitem [{\citenamefont {Sanchez}\ \emph {et~al.}(2012)\citenamefont
  {Sanchez}, \citenamefont {Chen}, \citenamefont {DeCamp}, \citenamefont
  {Heymann},\ and\ \citenamefont {Dogic}}]{Sanchez2012}%
  \BibitemOpen
  \bibfield  {author} {\bibinfo {author} {\bibfnamefont {T.}~\bibnamefont
  {Sanchez}}, \bibinfo {author} {\bibfnamefont {D.~T.~N.}\ \bibnamefont
  {Chen}}, \bibinfo {author} {\bibfnamefont {S.~J.}\ \bibnamefont {DeCamp}},
  \bibinfo {author} {\bibfnamefont {M.}~\bibnamefont {Heymann}},\ and\ \bibinfo
  {author} {\bibfnamefont {Z.}~\bibnamefont {Dogic}},\ }\bibfield  {title}
  {\bibinfo {title} {{Spontaneous motion in hierarchically assembled active
  matter}},\ }\href@noop {} {\bibfield  {journal} {\bibinfo  {journal}
  {Nature}\ }\textbf {\bibinfo {volume} {491}},\ \bibinfo {pages} {431}
  (\bibinfo {year} {2012})}\BibitemShut {NoStop}%
\bibitem [{\citenamefont {Howse}\ \emph {et~al.}(2007)\citenamefont {Howse},
  \citenamefont {Jones}, \citenamefont {Ryan}, \citenamefont {Gough},
  \citenamefont {Vafabakhsh},\ and\ \citenamefont
  {Golestanian}}]{Howse:2007hc}%
  \BibitemOpen
  \bibfield  {author} {\bibinfo {author} {\bibfnamefont {J.~R.}\ \bibnamefont
  {Howse}}, \bibinfo {author} {\bibfnamefont {R.~A.~L.}\ \bibnamefont {Jones}},
  \bibinfo {author} {\bibfnamefont {A.~J.}\ \bibnamefont {Ryan}}, \bibinfo
  {author} {\bibfnamefont {T.}~\bibnamefont {Gough}}, \bibinfo {author}
  {\bibfnamefont {R.}~\bibnamefont {Vafabakhsh}},\ and\ \bibinfo {author}
  {\bibfnamefont {R.}~\bibnamefont {Golestanian}},\ }\bibfield  {title}
  {\bibinfo {title} {{Self-Motile Colloidal Particles: From Directed Propulsion
  to Random Walk}},\ }\href@noop {} {\bibfield  {journal} {\bibinfo  {journal}
  {Physical Review Letters}\ }\textbf {\bibinfo {volume} {99}},\ \bibinfo
  {pages} {048102} (\bibinfo {year} {2007})}\BibitemShut {NoStop}%
\bibitem [{\citenamefont {Schuerle}\ \emph {et~al.}(2019)\citenamefont
  {Schuerle}, \citenamefont {Soleimany}, \citenamefont {Yeh}, \citenamefont
  {Anand}, \citenamefont {H{\"{a}}berli}, \citenamefont {Fleming},
  \citenamefont {Mirkhani}, \citenamefont {Qiu}, \citenamefont {Hauert},
  \citenamefont {Wang}, \citenamefont {Nelson},\ and\ \citenamefont
  {Bhatia}}]{Schuerle2019}%
  \BibitemOpen
  \bibfield  {author} {\bibinfo {author} {\bibfnamefont {S.}~\bibnamefont
  {Schuerle}}, \bibinfo {author} {\bibfnamefont {A.~P.}\ \bibnamefont
  {Soleimany}}, \bibinfo {author} {\bibfnamefont {T.}~\bibnamefont {Yeh}},
  \bibinfo {author} {\bibfnamefont {G.~M.}\ \bibnamefont {Anand}}, \bibinfo
  {author} {\bibfnamefont {M.}~\bibnamefont {H{\"{a}}berli}}, \bibinfo {author}
  {\bibfnamefont {H.~E.}\ \bibnamefont {Fleming}}, \bibinfo {author}
  {\bibfnamefont {N.}~\bibnamefont {Mirkhani}}, \bibinfo {author}
  {\bibfnamefont {F.}~\bibnamefont {Qiu}}, \bibinfo {author} {\bibfnamefont
  {S.}~\bibnamefont {Hauert}}, \bibinfo {author} {\bibfnamefont
  {X.}~\bibnamefont {Wang}}, \bibinfo {author} {\bibfnamefont {B.~J.}\
  \bibnamefont {Nelson}},\ and\ \bibinfo {author} {\bibfnamefont {S.~N.}\
  \bibnamefont {Bhatia}},\ }\bibfield  {title} {\bibinfo {title} {{Synthetic
  and living micropropellers for convection-enhanced nanoparticle transport}},\
  }\href@noop {} {\bibfield  {journal} {\bibinfo  {journal} {Science Advances}\
  }\textbf {\bibinfo {volume} {5}},\ \bibinfo {pages} {eaav4803} (\bibinfo
  {year} {2019})}\BibitemShut {NoStop}%
\bibitem [{\citenamefont {Meng}\ \emph {et~al.}(2021)\citenamefont {Meng},
  \citenamefont {Matsunaga}, \citenamefont {Mahault},\ and\ \citenamefont
  {Golestanian}}]{Meng2021}%
  \BibitemOpen
  \bibfield  {author} {\bibinfo {author} {\bibfnamefont {F.}~\bibnamefont
  {Meng}}, \bibinfo {author} {\bibfnamefont {D.}~\bibnamefont {Matsunaga}},
  \bibinfo {author} {\bibfnamefont {B.}~\bibnamefont {Mahault}},\ and\ \bibinfo
  {author} {\bibfnamefont {R.}~\bibnamefont {Golestanian}},\ }\bibfield
  {title} {\bibinfo {title} {{Magnetic Microswimmers Exhibit Bose-Einstein-Like
  Condensation}},\ }\href@noop {} {\bibfield  {journal} {\bibinfo  {journal}
  {Physical Review Letters}\ }\textbf {\bibinfo {volume} {126}},\ \bibinfo
  {pages} {078001} (\bibinfo {year} {2021})}\BibitemShut {NoStop}%
\bibitem [{\citenamefont {Zhou}\ \emph {et~al.}(2017)\citenamefont {Zhou},
  \citenamefont {Zhao}, \citenamefont {Wei},\ and\ \citenamefont
  {Wang}}]{Zhou2017}%
  \BibitemOpen
  \bibfield  {author} {\bibinfo {author} {\bibfnamefont {C.}~\bibnamefont
  {Zhou}}, \bibinfo {author} {\bibfnamefont {L.}~\bibnamefont {Zhao}}, \bibinfo
  {author} {\bibfnamefont {M.}~\bibnamefont {Wei}},\ and\ \bibinfo {author}
  {\bibfnamefont {W.}~\bibnamefont {Wang}},\ }\bibfield  {title} {\bibinfo
  {title} {{Twists and turns of orbiting and spinning metallic microparticles
  powered by megahertz ultrasound}},\ }\href@noop {} {\bibfield  {journal}
  {\bibinfo  {journal} {ACS nano}\ }\textbf {\bibinfo {volume} {11}},\ \bibinfo
  {pages} {12668} (\bibinfo {year} {2017})}\BibitemShut {NoStop}%
\bibitem [{\citenamefont {Dombrowski}\ \emph {et~al.}(2004)\citenamefont
  {Dombrowski}, \citenamefont {Cisneros}, \citenamefont {Chatkaew},
  \citenamefont {Goldstein},\ and\ \citenamefont {Kessler}}]{Dombrowski2004}%
  \BibitemOpen
  \bibfield  {author} {\bibinfo {author} {\bibfnamefont {C.}~\bibnamefont
  {Dombrowski}}, \bibinfo {author} {\bibfnamefont {L.}~\bibnamefont
  {Cisneros}}, \bibinfo {author} {\bibfnamefont {S.}~\bibnamefont {Chatkaew}},
  \bibinfo {author} {\bibfnamefont {R.~E.}\ \bibnamefont {Goldstein}},\ and\
  \bibinfo {author} {\bibfnamefont {J.~O.}\ \bibnamefont {Kessler}},\
  }\bibfield  {title} {\bibinfo {title} {{Self-Concentration and Large-Scale
  Coherence in Bacterial Dynamics}},\ }\href@noop {} {\bibfield  {journal}
  {\bibinfo  {journal} {Physical Review Letters}\ }\textbf {\bibinfo {volume}
  {93}},\ \bibinfo {pages} {098103} (\bibinfo {year} {2004})}\BibitemShut
  {NoStop}%
\bibitem [{\citenamefont {Drescher}\ \emph {et~al.}(2009)\citenamefont
  {Drescher}, \citenamefont {Leptos}, \citenamefont {Tuval}, \citenamefont
  {Ishikawa}, \citenamefont {Pedley},\ and\ \citenamefont
  {Goldstein}}]{Drescher:2009vn}%
  \BibitemOpen
  \bibfield  {author} {\bibinfo {author} {\bibfnamefont {K.}~\bibnamefont
  {Drescher}}, \bibinfo {author} {\bibfnamefont {K.~C.}\ \bibnamefont
  {Leptos}}, \bibinfo {author} {\bibfnamefont {I.}~\bibnamefont {Tuval}},
  \bibinfo {author} {\bibfnamefont {T.}~\bibnamefont {Ishikawa}}, \bibinfo
  {author} {\bibfnamefont {T.~J.}\ \bibnamefont {Pedley}},\ and\ \bibinfo
  {author} {\bibfnamefont {R.~E.}\ \bibnamefont {Goldstein}},\ }\bibfield
  {title} {\bibinfo {title} {{Dancing Volvox: Hydrodynamic Bound States of
  Swimming Algae}},\ }\href@noop {} {\bibfield  {journal} {\bibinfo  {journal}
  {Physical Review Letters}\ }\textbf {\bibinfo {volume} {102}},\ \bibinfo
  {pages} {168101} (\bibinfo {year} {2009})}\BibitemShut {NoStop}%
\bibitem [{\citenamefont {Thutupalli}\ \emph {et~al.}(2018)\citenamefont
  {Thutupalli}, \citenamefont {Geyer}, \citenamefont {Singh}, \citenamefont
  {Adhikari},\ and\ \citenamefont {Stone}}]{Thutupalli2018}%
  \BibitemOpen
  \bibfield  {author} {\bibinfo {author} {\bibfnamefont {S.}~\bibnamefont
  {Thutupalli}}, \bibinfo {author} {\bibfnamefont {D.}~\bibnamefont {Geyer}},
  \bibinfo {author} {\bibfnamefont {R.}~\bibnamefont {Singh}}, \bibinfo
  {author} {\bibfnamefont {R.}~\bibnamefont {Adhikari}},\ and\ \bibinfo
  {author} {\bibfnamefont {H.~A.}\ \bibnamefont {Stone}},\ }\bibfield  {title}
  {\bibinfo {title} {{Flow-induced phase separation of active particles is
  controlled by boundary conditions}},\ }\href@noop {} {\bibfield  {journal}
  {\bibinfo  {journal} {Proceedings of the National Academy of Sciences}\
  }\textbf {\bibinfo {volume} {115}},\ \bibinfo {pages} {5403} (\bibinfo {year}
  {2018})}\BibitemShut {NoStop}%
\bibitem [{\citenamefont {Driscoll}\ \emph {et~al.}(2017)\citenamefont
  {Driscoll}, \citenamefont {Delmotte}, \citenamefont {Youssef}, \citenamefont
  {Sacanna}, \citenamefont {Donev},\ and\ \citenamefont
  {Chaikin}}]{Driscoll2017}%
  \BibitemOpen
  \bibfield  {author} {\bibinfo {author} {\bibfnamefont {M.}~\bibnamefont
  {Driscoll}}, \bibinfo {author} {\bibfnamefont {B.}~\bibnamefont {Delmotte}},
  \bibinfo {author} {\bibfnamefont {M.}~\bibnamefont {Youssef}}, \bibinfo
  {author} {\bibfnamefont {S.}~\bibnamefont {Sacanna}}, \bibinfo {author}
  {\bibfnamefont {A.}~\bibnamefont {Donev}},\ and\ \bibinfo {author}
  {\bibfnamefont {P.}~\bibnamefont {Chaikin}},\ }\bibfield  {title} {\bibinfo
  {title} {{Unstable fronts and motile structures formed by microrollers}},\
  }\href@noop {} {\bibfield  {journal} {\bibinfo  {journal} {Nature Physics}\
  }\textbf {\bibinfo {volume} {13}},\ \bibinfo {pages} {375} (\bibinfo {year}
  {2017})}\BibitemShut {NoStop}%
\bibitem [{\citenamefont {Delmotte}(2019)}]{Delmotte2019}%
  \BibitemOpen
  \bibfield  {author} {\bibinfo {author} {\bibfnamefont {B.}~\bibnamefont
  {Delmotte}},\ }\bibfield  {title} {\bibinfo {title} {{Hydrodynamically bound
  states of a pair of microrollers: A dynamical system insight}},\ }\href@noop
  {} {\bibfield  {journal} {\bibinfo  {journal} {Physical Review Fluids}\
  }\textbf {\bibinfo {volume} {4}},\ \bibinfo {pages} {044302} (\bibinfo {year}
  {2019})}\BibitemShut {NoStop}%
\bibitem [{\citenamefont {Pedley}\ and\ \citenamefont
  {Kessler}(1990)}]{Pedley1990}%
  \BibitemOpen
  \bibfield  {author} {\bibinfo {author} {\bibfnamefont {T.~J.}\ \bibnamefont
  {Pedley}}\ and\ \bibinfo {author} {\bibfnamefont {J.~O.}\ \bibnamefont
  {Kessler}},\ }\bibfield  {title} {\bibinfo {title} {{A new continuum model
  for suspensions of gyrotactic micro-organisms.}},\ }\href@noop {} {\bibfield
  {journal} {\bibinfo  {journal} {Journal of Fluid Mechanics}\ }\textbf
  {\bibinfo {volume} {212}},\ \bibinfo {pages} {155} (\bibinfo {year}
  {1990})}\BibitemShut {NoStop}%
\bibitem [{\citenamefont {Wensink}\ \emph {et~al.}(2012)\citenamefont
  {Wensink}, \citenamefont {Dunkel}, \citenamefont {Heidenreich}, \citenamefont
  {Drescher}, \citenamefont {Goldstein}, \citenamefont {L{\"{o}}wen},\ and\
  \citenamefont {Yeomans}}]{Wensink:2012}%
  \BibitemOpen
  \bibfield  {author} {\bibinfo {author} {\bibfnamefont {H.~H.}\ \bibnamefont
  {Wensink}}, \bibinfo {author} {\bibfnamefont {J.}~\bibnamefont {Dunkel}},
  \bibinfo {author} {\bibfnamefont {S.}~\bibnamefont {Heidenreich}}, \bibinfo
  {author} {\bibfnamefont {K.}~\bibnamefont {Drescher}}, \bibinfo {author}
  {\bibfnamefont {R.~E.}\ \bibnamefont {Goldstein}}, \bibinfo {author}
  {\bibfnamefont {H.}~\bibnamefont {L{\"{o}}wen}},\ and\ \bibinfo {author}
  {\bibfnamefont {J.~M.}\ \bibnamefont {Yeomans}},\ }\bibfield  {title}
  {\bibinfo {title} {{Meso-scale turbulence in living fluids}},\ }\href@noop {}
  {\bibfield  {journal} {\bibinfo  {journal} {Proceedings of the National
  Academy of Sciences}\ }\textbf {\bibinfo {volume} {109}},\ \bibinfo {pages}
  {14308} (\bibinfo {year} {2012})}\BibitemShut {NoStop}%
\bibitem [{\citenamefont {Ishikawa}\ \emph {et~al.}(2006)\citenamefont
  {Ishikawa}, \citenamefont {Simmonds},\ and\ \citenamefont
  {Pedley}}]{Ishikawa2006}%
  \BibitemOpen
  \bibfield  {author} {\bibinfo {author} {\bibfnamefont {T.}~\bibnamefont
  {Ishikawa}}, \bibinfo {author} {\bibfnamefont {M.~P.}\ \bibnamefont
  {Simmonds}},\ and\ \bibinfo {author} {\bibfnamefont {T.~J.}\ \bibnamefont
  {Pedley}},\ }\bibfield  {title} {\bibinfo {title} {{Hydrodynamic interaction
  of two swimming model micro-organisms}},\ }\href@noop {} {\bibfield
  {journal} {\bibinfo  {journal} {Journal of Fluid Mechanics}\ }\textbf
  {\bibinfo {volume} {568}},\ \bibinfo {pages} {119} (\bibinfo {year}
  {2006})}\BibitemShut {NoStop}%
\bibitem [{\citenamefont {Ishikawa}\ \emph {et~al.}(2008)\citenamefont
  {Ishikawa}, \citenamefont {Locsei},\ and\ \citenamefont
  {Pedley}}]{Ishikawa2008}%
  \BibitemOpen
  \bibfield  {author} {\bibinfo {author} {\bibfnamefont {T.}~\bibnamefont
  {Ishikawa}}, \bibinfo {author} {\bibfnamefont {J.~T.}\ \bibnamefont
  {Locsei}},\ and\ \bibinfo {author} {\bibfnamefont {T.~J.}\ \bibnamefont
  {Pedley}},\ }\bibfield  {title} {\bibinfo {title} {{Development of coherent
  structures in concentrated suspensions of swimming model micro-organisms}},\
  }\href@noop {} {\bibfield  {journal} {\bibinfo  {journal} {Journal of Fluid
  Mechanics}\ }\textbf {\bibinfo {volume} {615}},\ \bibinfo {pages} {401}
  (\bibinfo {year} {2008})}\BibitemShut {NoStop}%
\bibitem [{\citenamefont {Ishikawa}\ \emph {et~al.}(2021)\citenamefont
  {Ishikawa}, \citenamefont {Brumley},\ and\ \citenamefont
  {Pedley}}]{ishikawa2021}%
  \BibitemOpen
  \bibfield  {author} {\bibinfo {author} {\bibfnamefont {T.}~\bibnamefont
  {Ishikawa}}, \bibinfo {author} {\bibfnamefont {D.~R.}\ \bibnamefont
  {Brumley}},\ and\ \bibinfo {author} {\bibfnamefont {T.~J.}\ \bibnamefont
  {Pedley}},\ }\bibfield  {title} {\bibinfo {title} {{Rheology of a
  concentrated suspension of spherical squirmers: monolayer in simple shear
  flow}},\ }\href@noop {} {\bibfield  {journal} {\bibinfo  {journal} {Journal
  of Fluid Mechanics}\ }\textbf {\bibinfo {volume} {914}},\ \bibinfo {pages}
  {A26} (\bibinfo {year} {2021})}\BibitemShut {NoStop}%
\bibitem [{\citenamefont {Brumley}\ and\ \citenamefont
  {Pedley}(2019)}]{BrumleyPRF2019}%
  \BibitemOpen
  \bibfield  {author} {\bibinfo {author} {\bibfnamefont {D.~R.}\ \bibnamefont
  {Brumley}}\ and\ \bibinfo {author} {\bibfnamefont {T.~J.}\ \bibnamefont
  {Pedley}},\ }\bibfield  {title} {\bibinfo {title} {{Stability of arrays of
  bottom-heavy spherical squirmers}},\ }\href@noop {} {\bibfield  {journal}
  {\bibinfo  {journal} {Physical Review Fluids}\ }\textbf {\bibinfo {volume}
  {4}},\ \bibinfo {pages} {053102} (\bibinfo {year} {2019})}\BibitemShut
  {NoStop}%
\bibitem [{\citenamefont {Wei}\ \emph {et~al.}(2021)\citenamefont {Wei},
  \citenamefont {Dehnavi}, \citenamefont {Aubin-Tam},\ and\ \citenamefont
  {Tam}}]{Wei2021}%
  \BibitemOpen
  \bibfield  {author} {\bibinfo {author} {\bibfnamefont {D.}~\bibnamefont
  {Wei}}, \bibinfo {author} {\bibfnamefont {P.~G.}\ \bibnamefont {Dehnavi}},
  \bibinfo {author} {\bibfnamefont {M.-E.}\ \bibnamefont {Aubin-Tam}},\ and\
  \bibinfo {author} {\bibfnamefont {D.}~\bibnamefont {Tam}},\ }\bibfield
  {title} {\bibinfo {title} {{Measurements of the unsteady flow field around
  beating cilia}},\ }\href@noop {} {\bibfield  {journal} {\bibinfo  {journal}
  {Journal of Fluid Mechanics}\ }\textbf {\bibinfo {volume} {915}} (\bibinfo
  {year} {2021})}\BibitemShut {NoStop}%
\bibitem [{\citenamefont {Brumley}\ \emph {et~al.}(2014)\citenamefont
  {Brumley}, \citenamefont {Wan}, \citenamefont {Polin},\ and\ \citenamefont
  {Goldstein}}]{Brumley2014}%
  \BibitemOpen
  \bibfield  {author} {\bibinfo {author} {\bibfnamefont {D.~R.}\ \bibnamefont
  {Brumley}}, \bibinfo {author} {\bibfnamefont {K.~Y.}\ \bibnamefont {Wan}},
  \bibinfo {author} {\bibfnamefont {M.}~\bibnamefont {Polin}},\ and\ \bibinfo
  {author} {\bibfnamefont {R.~E.}\ \bibnamefont {Goldstein}},\ }\bibfield
  {title} {\bibinfo {title} {{Flagellar synchronization through direct
  hydrodynamic interactions}},\ }\href@noop {} {\bibfield  {journal} {\bibinfo
  {journal} {{eLife}}\ }\textbf {\bibinfo {volume} {3}},\ \bibinfo {pages}
  {e02750} (\bibinfo {year} {2014})}\BibitemShut {NoStop}%
\bibitem [{\citenamefont {Brumley}\ \emph {et~al.}(2015)\citenamefont
  {Brumley}, \citenamefont {Rusconi}, \citenamefont {Son},\ and\ \citenamefont
  {Stocker}}]{Brumley2015}%
  \BibitemOpen
  \bibfield  {author} {\bibinfo {author} {\bibfnamefont {D.~R.}\ \bibnamefont
  {Brumley}}, \bibinfo {author} {\bibfnamefont {R.}~\bibnamefont {Rusconi}},
  \bibinfo {author} {\bibfnamefont {K.}~\bibnamefont {Son}},\ and\ \bibinfo
  {author} {\bibfnamefont {R.}~\bibnamefont {Stocker}},\ }\bibfield  {title}
  {\bibinfo {title} {{Flagella, flexibility and flow: Physical processes in
  microbial ecology}},\ }\href@noop {} {\bibfield  {journal} {\bibinfo
  {journal} {The European Physical Journal Special Topics}\ }\textbf {\bibinfo
  {volume} {224}},\ \bibinfo {pages} {3119} (\bibinfo {year}
  {2015})}\BibitemShut {NoStop}%
\bibitem [{\citenamefont {Drescher}\ \emph {et~al.}(2010)\citenamefont
  {Drescher}, \citenamefont {Goldstein}, \citenamefont {Michel}, \citenamefont
  {Polin},\ and\ \citenamefont {Tuval}}]{Drescher:2010kx}%
  \BibitemOpen
  \bibfield  {author} {\bibinfo {author} {\bibfnamefont {K.}~\bibnamefont
  {Drescher}}, \bibinfo {author} {\bibfnamefont {R.~E.}\ \bibnamefont
  {Goldstein}}, \bibinfo {author} {\bibfnamefont {N.}~\bibnamefont {Michel}},
  \bibinfo {author} {\bibfnamefont {M.}~\bibnamefont {Polin}},\ and\ \bibinfo
  {author} {\bibfnamefont {I.}~\bibnamefont {Tuval}},\ }\bibfield  {title}
  {\bibinfo {title} {{Direct Measurement of the Flow Field around Swimming
  Microorganisms}},\ }\href@noop {} {\bibfield  {journal} {\bibinfo  {journal}
  {Physical Review Letters}\ }\textbf {\bibinfo {volume} {105}},\ \bibinfo
  {pages} {168101} (\bibinfo {year} {2010})}\BibitemShut {NoStop}%
\bibitem [{\citenamefont {Guasto}\ \emph {et~al.}(2010)\citenamefont {Guasto},
  \citenamefont {Johnson},\ and\ \citenamefont {Gollub}}]{Guasto:2010ly}%
  \BibitemOpen
  \bibfield  {author} {\bibinfo {author} {\bibfnamefont {J.~S.}\ \bibnamefont
  {Guasto}}, \bibinfo {author} {\bibfnamefont {K.~A.}\ \bibnamefont
  {Johnson}},\ and\ \bibinfo {author} {\bibfnamefont {J.~P.}\ \bibnamefont
  {Gollub}},\ }\bibfield  {title} {\bibinfo {title} {{Oscillatory Flows Induced
  by Microorganisms Swimming in Two Dimensions}},\ }\href@noop {} {\bibfield
  {journal} {\bibinfo  {journal} {Physical Review Letters}\ }\textbf {\bibinfo
  {volume} {105}},\ \bibinfo {pages} {168102} (\bibinfo {year}
  {2010})}\BibitemShut {NoStop}%
\bibitem [{\citenamefont {Drescher}\ \emph {et~al.}(2011)\citenamefont
  {Drescher}, \citenamefont {Dunkel}, \citenamefont {Cisneros}, \citenamefont
  {Ganguly},\ and\ \citenamefont {Goldstein}}]{Drescher:2011bh}%
  \BibitemOpen
  \bibfield  {author} {\bibinfo {author} {\bibfnamefont {K.}~\bibnamefont
  {Drescher}}, \bibinfo {author} {\bibfnamefont {J.}~\bibnamefont {Dunkel}},
  \bibinfo {author} {\bibfnamefont {L.~H.}\ \bibnamefont {Cisneros}}, \bibinfo
  {author} {\bibfnamefont {S.}~\bibnamefont {Ganguly}},\ and\ \bibinfo {author}
  {\bibfnamefont {R.~E.}\ \bibnamefont {Goldstein}},\ }\bibfield  {title}
  {\bibinfo {title} {{Fluid dynamics and noise in bacterial cell--cell and
  cell--surface scattering}},\ }\href@noop {} {\bibfield  {journal} {\bibinfo
  {journal} {Proceedings of the National Academy of Sciences}\ }\textbf
  {\bibinfo {volume} {108}},\ \bibinfo {pages} {10940} (\bibinfo {year}
  {2011})}\BibitemShut {NoStop}%
\bibitem [{\citenamefont {Cortese}\ and\ \citenamefont
  {Wan}(2021)}]{Cortese2021}%
  \BibitemOpen
  \bibfield  {author} {\bibinfo {author} {\bibfnamefont {D.}~\bibnamefont
  {Cortese}}\ and\ \bibinfo {author} {\bibfnamefont {K.~Y.}\ \bibnamefont
  {Wan}},\ }\bibfield  {title} {\bibinfo {title} {{Control of Helical
  Navigation by Three-Dimensional Flagellar Beating}},\ }\href@noop {}
  {\bibfield  {journal} {\bibinfo  {journal} {Physical Review Letters}\
  }\textbf {\bibinfo {volume} {126}},\ \bibinfo {pages} {088003} (\bibinfo
  {year} {2021})}\BibitemShut {NoStop}%
\bibitem [{\citenamefont {Lippera}\ \emph {et~al.}(2021)\citenamefont
  {Lippera}, \citenamefont {Benzaquen},\ and\ \citenamefont
  {Michelin}}]{Lippera2021}%
  \BibitemOpen
  \bibfield  {author} {\bibinfo {author} {\bibfnamefont {K.}~\bibnamefont
  {Lippera}}, \bibinfo {author} {\bibfnamefont {M.}~\bibnamefont {Benzaquen}},\
  and\ \bibinfo {author} {\bibfnamefont {S.}~\bibnamefont {Michelin}},\
  }\bibfield  {title} {\bibinfo {title} {{Alignment and scattering of colliding
  active droplets}},\ }\href@noop {} {\bibfield  {journal} {\bibinfo  {journal}
  {Soft Matter}\ }\textbf {\bibinfo {volume} {17}},\ \bibinfo {pages} {365}
  (\bibinfo {year} {2021})}\BibitemShut {NoStop}%
\bibitem [{\citenamefont {Dehkharghani}\ \emph {et~al.}(2019)\citenamefont
  {Dehkharghani}, \citenamefont {Waisbord}, \citenamefont {Dunkel},\ and\
  \citenamefont {Guasto}}]{Dehkharghani2019}%
  \BibitemOpen
  \bibfield  {author} {\bibinfo {author} {\bibfnamefont {A.}~\bibnamefont
  {Dehkharghani}}, \bibinfo {author} {\bibfnamefont {N.}~\bibnamefont
  {Waisbord}}, \bibinfo {author} {\bibfnamefont {J.}~\bibnamefont {Dunkel}},\
  and\ \bibinfo {author} {\bibfnamefont {J.~S.}\ \bibnamefont {Guasto}},\
  }\bibfield  {title} {\bibinfo {title} {{Bacterial scattering in microfluidic
  crystal flows reveals giant active Taylor--Aris dispersion}},\ }\href@noop {}
  {\bibfield  {journal} {\bibinfo  {journal} {Proceedings of the National
  Academy of Sciences}\ }\textbf {\bibinfo {volume} {116}},\ \bibinfo {pages}
  {11119} (\bibinfo {year} {2019})}\BibitemShut {NoStop}%
\bibitem [{\citenamefont {Lauga}\ \emph {et~al.}(2006)\citenamefont {Lauga},
  \citenamefont {DiLuzio}, \citenamefont {Whitesides},\ and\ \citenamefont
  {Stone}}]{Lauga2006}%
  \BibitemOpen
  \bibfield  {author} {\bibinfo {author} {\bibfnamefont {E.}~\bibnamefont
  {Lauga}}, \bibinfo {author} {\bibfnamefont {W.~R.}\ \bibnamefont {DiLuzio}},
  \bibinfo {author} {\bibfnamefont {G.~M.}\ \bibnamefont {Whitesides}},\ and\
  \bibinfo {author} {\bibfnamefont {H.~A.}\ \bibnamefont {Stone}},\ }\bibfield
  {title} {\bibinfo {title} {{Swimming in circles: motion of bacteria near
  solid boundaries}},\ }\href@noop {} {\bibfield  {journal} {\bibinfo
  {journal} {{Biophysical Journal}}\ }\textbf {\bibinfo {volume} {90}},\
  \bibinfo {pages} {400} (\bibinfo {year} {2006})}\BibitemShut {NoStop}%
\bibitem [{\citenamefont {Hoeger}\ and\ \citenamefont
  {Ursell}(2021)}]{Hoeger2021}%
  \BibitemOpen
  \bibfield  {author} {\bibinfo {author} {\bibfnamefont {K.}~\bibnamefont
  {Hoeger}}\ and\ \bibinfo {author} {\bibfnamefont {T.}~\bibnamefont
  {Ursell}},\ }\bibfield  {title} {\bibinfo {title} {{Steric scattering of
  rod-like swimmers in low Reynolds number environments}},\ }\href@noop {}
  {\bibfield  {journal} {\bibinfo  {journal} {Soft Matter}\ }\textbf {\bibinfo
  {volume} {17}},\ \bibinfo {pages} {2479} (\bibinfo {year}
  {2021})}\BibitemShut {NoStop}%
\bibitem [{\citenamefont {Ishikawa}\ \emph {et~al.}(2020)\citenamefont
  {Ishikawa}, \citenamefont {Pedley}, \citenamefont {Drescher},\ and\
  \citenamefont {Goldstein}}]{Ishikawa2020}%
  \BibitemOpen
  \bibfield  {author} {\bibinfo {author} {\bibfnamefont {T.}~\bibnamefont
  {Ishikawa}}, \bibinfo {author} {\bibfnamefont {T.~J.}\ \bibnamefont
  {Pedley}}, \bibinfo {author} {\bibfnamefont {K.}~\bibnamefont {Drescher}},\
  and\ \bibinfo {author} {\bibfnamefont {R.~E.}\ \bibnamefont {Goldstein}},\
  }\bibfield  {title} {\bibinfo {title} {{Stability of dancing Volvox}},\
  }\href@noop {} {\bibfield  {journal} {\bibinfo  {journal} {Journal of Fluid
  Mechanics}\ }\textbf {\bibinfo {volume} {903}},\ \bibinfo {pages} {A11}
  (\bibinfo {year} {2020})}\BibitemShut {NoStop}%
\bibitem [{\citenamefont {Contino}\ \emph {et~al.}(2015)\citenamefont
  {Contino}, \citenamefont {Lushi}, \citenamefont {Tuval}, \citenamefont
  {Kantsler},\ and\ \citenamefont {Polin}}]{Contino2015}%
  \BibitemOpen
  \bibfield  {author} {\bibinfo {author} {\bibfnamefont {M.}~\bibnamefont
  {Contino}}, \bibinfo {author} {\bibfnamefont {E.}~\bibnamefont {Lushi}},
  \bibinfo {author} {\bibfnamefont {I.}~\bibnamefont {Tuval}}, \bibinfo
  {author} {\bibfnamefont {V.}~\bibnamefont {Kantsler}},\ and\ \bibinfo
  {author} {\bibfnamefont {M.}~\bibnamefont {Polin}},\ }\bibfield  {title}
  {\bibinfo {title} {{Microalgae Scatter off Solid Surfaces by Hydrodynamic and
  Contact Forces}},\ }\href@noop {} {\bibfield  {journal} {\bibinfo  {journal}
  {Physical Review Letters}\ }\textbf {\bibinfo {volume} {115}},\ \bibinfo
  {pages} {258102} (\bibinfo {year} {2015})}\BibitemShut {NoStop}%
\bibitem [{\citenamefont {G{\"{o}}tze}\ and\ \citenamefont
  {Gompper}(2010)}]{Gotze2010}%
  \BibitemOpen
  \bibfield  {author} {\bibinfo {author} {\bibfnamefont {I.~O.}\ \bibnamefont
  {G{\"{o}}tze}}\ and\ \bibinfo {author} {\bibfnamefont {G.}~\bibnamefont
  {Gompper}},\ }\bibfield  {title} {\bibinfo {title} {{Mesoscale simulations of
  hydrodynamic squirmer interactions}},\ }\href@noop {} {\bibfield  {journal}
  {\bibinfo  {journal} {Physical Review E}\ }\textbf {\bibinfo {volume} {82}},\
  \bibinfo {pages} {041921} (\bibinfo {year} {2010})}\BibitemShut {NoStop}%
\bibitem [{\citenamefont {Llopis}\ and\ \citenamefont
  {Pagonabarraga}(2010)}]{Llopis:2010}%
  \BibitemOpen
  \bibfield  {author} {\bibinfo {author} {\bibfnamefont {I.}~\bibnamefont
  {Llopis}}\ and\ \bibinfo {author} {\bibfnamefont {I.}~\bibnamefont
  {Pagonabarraga}},\ }\bibfield  {title} {\bibinfo {title} {{Hydrodynamic
  interactions in squirmer motion: Swimming with a neighbour and close to a
  wall}},\ }\href@noop {} {\bibfield  {journal} {\bibinfo  {journal} {Journal
  of Non-Newtonian Fluid Mechanics}\ }\textbf {\bibinfo {volume} {165}},\
  \bibinfo {pages} {946} (\bibinfo {year} {2010})}\BibitemShut {NoStop}%
\bibitem [{\citenamefont {Li}\ \emph {et~al.}(2016)\citenamefont {Li},
  \citenamefont {Ostace},\ and\ \citenamefont {Ardekani}}]{Li2016}%
  \BibitemOpen
  \bibfield  {author} {\bibinfo {author} {\bibfnamefont {G.}~\bibnamefont
  {Li}}, \bibinfo {author} {\bibfnamefont {A.}~\bibnamefont {Ostace}},\ and\
  \bibinfo {author} {\bibfnamefont {A.~M.}\ \bibnamefont {Ardekani}},\
  }\bibfield  {title} {\bibinfo {title} {{Hydrodynamic interaction of swimming
  organisms in an inertial regime}},\ }\href@noop {} {\bibfield  {journal}
  {\bibinfo  {journal} {Physical Review E}\ }\textbf {\bibinfo {volume} {94}},\
  \bibinfo {pages} {053104} (\bibinfo {year} {2016})}\BibitemShut {NoStop}%
\bibitem [{\citenamefont {More}\ and\ \citenamefont
  {Ardekani}(2021)}]{More2021}%
  \BibitemOpen
  \bibfield  {author} {\bibinfo {author} {\bibfnamefont {R.~V.}\ \bibnamefont
  {More}}\ and\ \bibinfo {author} {\bibfnamefont {A.~M.}\ \bibnamefont
  {Ardekani}},\ }\bibfield  {title} {\bibinfo {title} {{Hydrodynamic
  interactions between swimming microorganisms in a linearly density stratified
  fluid}},\ }\href@noop {} {\bibfield  {journal} {\bibinfo  {journal} {Physical
  Review E}\ }\textbf {\bibinfo {volume} {103}},\ \bibinfo {pages} {013109}
  (\bibinfo {year} {2021})}\BibitemShut {NoStop}%
\bibitem [{\citenamefont {Giacch{\'{e}}}\ and\ \citenamefont
  {Ishikawa}(2010)}]{Giacche2010}%
  \BibitemOpen
  \bibfield  {author} {\bibinfo {author} {\bibfnamefont {D.}~\bibnamefont
  {Giacch{\'{e}}}}\ and\ \bibinfo {author} {\bibfnamefont {T.}~\bibnamefont
  {Ishikawa}},\ }\bibfield  {title} {\bibinfo {title} {{Hydrodynamic
  interaction of two unsteady model microorganisms}},\ }\href@noop {}
  {\bibfield  {journal} {\bibinfo  {journal} {Journal of Theoretical Biology}\
  }\textbf {\bibinfo {volume} {267}},\ \bibinfo {pages} {252} (\bibinfo {year}
  {2010})}\BibitemShut {NoStop}%
\bibitem [{\citenamefont {Papavassiliou}\ and\ \citenamefont
  {Alexander}(2017)}]{Papavassiliou2017}%
  \BibitemOpen
  \bibfield  {author} {\bibinfo {author} {\bibfnamefont {D.}~\bibnamefont
  {Papavassiliou}}\ and\ \bibinfo {author} {\bibfnamefont {G.~P.}\ \bibnamefont
  {Alexander}},\ }\bibfield  {title} {\bibinfo {title} {{Exact solutions for
  hydrodynamic interactions of two squirming spheres}},\ }\href@noop {}
  {\bibfield  {journal} {\bibinfo  {journal} {Journal of Fluid Mechanics}\
  }\textbf {\bibinfo {volume} {813}},\ \bibinfo {pages} {618} (\bibinfo {year}
  {2017})}\BibitemShut {NoStop}%
\bibitem [{\citenamefont {Blake}(1971)}]{Blake:Squirmer}%
  \BibitemOpen
  \bibfield  {author} {\bibinfo {author} {\bibfnamefont {J.~R.}\ \bibnamefont
  {Blake}},\ }\bibfield  {title} {\bibinfo {title} {{A spherical envelope
  approach to ciliary propulsion}},\ }\href@noop {} {\bibfield  {journal}
  {\bibinfo  {journal} {Journal of Fluid Mechanics}\ }\textbf {\bibinfo
  {volume} {46}},\ \bibinfo {pages} {199} (\bibinfo {year} {1971})}\BibitemShut
  {NoStop}%
\bibitem [{\citenamefont {Pedley}\ \emph {et~al.}(2016)\citenamefont {Pedley},
  \citenamefont {Brumley},\ and\ \citenamefont {Goldstein}}]{Pedley2016}%
  \BibitemOpen
  \bibfield  {author} {\bibinfo {author} {\bibfnamefont {T.~J.}\ \bibnamefont
  {Pedley}}, \bibinfo {author} {\bibfnamefont {D.~R.}\ \bibnamefont
  {Brumley}},\ and\ \bibinfo {author} {\bibfnamefont {R.~E.}\ \bibnamefont
  {Goldstein}},\ }\bibfield  {title} {\bibinfo {title} {{Squirmers with swirl:
  a model for Volvox swimming}},\ }\href@noop {} {\bibfield  {journal}
  {\bibinfo  {journal} {Journal of Fluid Mechanics}\ }\textbf {\bibinfo
  {volume} {798}},\ \bibinfo {pages} {165} (\bibinfo {year}
  {2016})}\BibitemShut {NoStop}%
\bibitem [{\citenamefont {Kim}\ and\ \citenamefont
  {Karrila}(2005)}]{KimAndKarrila:Microhydrodynamics}%
  \BibitemOpen
  \bibfield  {author} {\bibinfo {author} {\bibfnamefont {S.}~\bibnamefont
  {Kim}}\ and\ \bibinfo {author} {\bibfnamefont {S.~J.}\ \bibnamefont
  {Karrila}},\ }\href@noop {} {\emph {\bibinfo {title} {{Microhydrodynamics -
  Principles and Selected Applications}}}}\ (\bibinfo  {publisher} {Dover
  Publications},\ \bibinfo {year} {2005})\BibitemShut {NoStop}%
\bibitem [{\citenamefont {Kessler}(1985)}]{Kessler:1985a}%
  \BibitemOpen
  \bibfield  {author} {\bibinfo {author} {\bibfnamefont {J.~O.}\ \bibnamefont
  {Kessler}},\ }\bibfield  {title} {\bibinfo {title} {{Hydrodynamic focusing of
  motile algal cells}},\ }\href@noop {} {\bibfield  {journal} {\bibinfo
  {journal} {Nature}\ }\textbf {\bibinfo {volume} {313}},\ \bibinfo {pages}
  {218} (\bibinfo {year} {1985})}\BibitemShut {NoStop}%
\bibitem [{\citenamefont {Sengupta}\ \emph {et~al.}(2017)\citenamefont
  {Sengupta}, \citenamefont {Carrara},\ and\ \citenamefont
  {Stocker}}]{Sengupta2017}%
  \BibitemOpen
  \bibfield  {author} {\bibinfo {author} {\bibfnamefont {A.}~\bibnamefont
  {Sengupta}}, \bibinfo {author} {\bibfnamefont {F.}~\bibnamefont {Carrara}},\
  and\ \bibinfo {author} {\bibfnamefont {R.}~\bibnamefont {Stocker}},\
  }\bibfield  {title} {\bibinfo {title} {{Phytoplankton can actively diversify
  their migration strategy in response to turbulent cues}},\ }\href@noop {}
  {\bibfield  {journal} {\bibinfo  {journal} {Nature}\ }\textbf {\bibinfo
  {volume} {543}},\ \bibinfo {pages} {555} (\bibinfo {year}
  {2017})}\BibitemShut {NoStop}%
\bibitem [{\citenamefont {Kyoya}\ \emph {et~al.}(2015)\citenamefont {Kyoya},
  \citenamefont {Matsunaga}, \citenamefont {Imai}, \citenamefont {Omori},\ and\
  \citenamefont {Ishikawa}}]{Kyoya2015}%
  \BibitemOpen
  \bibfield  {author} {\bibinfo {author} {\bibfnamefont {K.}~\bibnamefont
  {Kyoya}}, \bibinfo {author} {\bibfnamefont {D.}~\bibnamefont {Matsunaga}},
  \bibinfo {author} {\bibfnamefont {Y.}~\bibnamefont {Imai}}, \bibinfo {author}
  {\bibfnamefont {T.}~\bibnamefont {Omori}},\ and\ \bibinfo {author}
  {\bibfnamefont {T.}~\bibnamefont {Ishikawa}},\ }\bibfield  {title} {\bibinfo
  {title} {{Shape matters: Near-field fluid mechanics dominate the collective
  motions of ellipsoidal squirmers}},\ }\href@noop {} {\bibfield  {journal}
  {\bibinfo  {journal} {Physical Review E}\ }\textbf {\bibinfo {volume} {92}},\
  \bibinfo {pages} {063027} (\bibinfo {year} {2015})}\BibitemShut {NoStop}%
\bibitem [{\citenamefont {Lintuvuori}\ \emph {et~al.}(2016)\citenamefont
  {Lintuvuori}, \citenamefont {Brown}, \citenamefont {Stratford},\ and\
  \citenamefont {Marenduzzo}}]{Lintuvuori2016}%
  \BibitemOpen
  \bibfield  {author} {\bibinfo {author} {\bibfnamefont {J.~S.}\ \bibnamefont
  {Lintuvuori}}, \bibinfo {author} {\bibfnamefont {A.~T.}\ \bibnamefont
  {Brown}}, \bibinfo {author} {\bibfnamefont {K.}~\bibnamefont {Stratford}},\
  and\ \bibinfo {author} {\bibfnamefont {D.}~\bibnamefont {Marenduzzo}},\
  }\bibfield  {title} {\bibinfo {title} {{Hydrodynamic oscillations and
  variable swimming speed in squirmers close to repulsive walls}},\ }\href@noop
  {} {\bibfield  {journal} {\bibinfo  {journal} {Soft Matter}\ }\textbf
  {\bibinfo {volume} {12}},\ \bibinfo {pages} {7959} (\bibinfo {year}
  {2016})}\BibitemShut {NoStop}%
\bibitem [{\citenamefont {{K. V. S.}}\ and\ \citenamefont
  {Thampi}(2021)}]{Chaithanya2021}%
  \BibitemOpen
  \bibfield  {author} {\bibinfo {author} {\bibfnamefont {C.}~\bibnamefont {{K.
  V. S.}}}\ and\ \bibinfo {author} {\bibfnamefont {S.~P.}\ \bibnamefont
  {Thampi}},\ }\bibfield  {title} {\bibinfo {title} {{Wall-curvature driven
  dynamics of a microswimmer}},\ }\href@noop {} {\bibfield  {journal} {\bibinfo
   {journal} {Physical Review Fluids}\ }\textbf {\bibinfo {volume} {6}},\
  \bibinfo {pages} {083101} (\bibinfo {year} {2021})}\BibitemShut {NoStop}%
\bibitem [{\citenamefont {Varma}\ and\ \citenamefont
  {Michelin}(2019)}]{Varma2019}%
  \BibitemOpen
  \bibfield  {author} {\bibinfo {author} {\bibfnamefont {A.}~\bibnamefont
  {Varma}}\ and\ \bibinfo {author} {\bibfnamefont {S.}~\bibnamefont
  {Michelin}},\ }\bibfield  {title} {\bibinfo {title} {{Modeling
  chemo-hydrodynamic interactions of phoretic particles: A unified
  framework}},\ }\href@noop {} {\bibfield  {journal} {\bibinfo  {journal}
  {Physical Review Fluids}\ }\textbf {\bibinfo {volume} {4}},\ \bibinfo {pages}
  {124204} (\bibinfo {year} {2019})}\BibitemShut {NoStop}%
\bibitem [{\citenamefont {B{\'{a}}rdfalvy}\ \emph {et~al.}(2020)\citenamefont
  {B{\'{a}}rdfalvy}, \citenamefont {Anjum}, \citenamefont {Nardini},
  \citenamefont {Morozov},\ and\ \citenamefont {Stenhammar}}]{Bardfalvy2020}%
  \BibitemOpen
  \bibfield  {author} {\bibinfo {author} {\bibfnamefont {D.}~\bibnamefont
  {B{\'{a}}rdfalvy}}, \bibinfo {author} {\bibfnamefont {S.}~\bibnamefont
  {Anjum}}, \bibinfo {author} {\bibfnamefont {C.}~\bibnamefont {Nardini}},
  \bibinfo {author} {\bibfnamefont {A.}~\bibnamefont {Morozov}},\ and\ \bibinfo
  {author} {\bibfnamefont {J.}~\bibnamefont {Stenhammar}},\ }\bibfield  {title}
  {\bibinfo {title} {{Symmetric Mixtures of Pusher and Puller Microswimmers
  Behave as Noninteracting Suspensions}},\ }\href@noop {} {\bibfield  {journal}
  {\bibinfo  {journal} {Physical Review Letters}\ }\textbf {\bibinfo {volume}
  {125}},\ \bibinfo {pages} {018003} (\bibinfo {year} {2020})}\BibitemShut
  {NoStop}%
\bibitem [{\citenamefont {Nassios}\ \emph {et~al.}(2016)\citenamefont
  {Nassios}, \citenamefont {Yap},\ and\ \citenamefont {Sader}}]{Nassios2016}%
  \BibitemOpen
  \bibfield  {author} {\bibinfo {author} {\bibfnamefont {J.}~\bibnamefont
  {Nassios}}, \bibinfo {author} {\bibfnamefont {Y.~W.}\ \bibnamefont {Yap}},\
  and\ \bibinfo {author} {\bibfnamefont {J.~E.}\ \bibnamefont {Sader}},\
  }\bibfield  {title} {\bibinfo {title} {{Flow generated by oscillatory uniform
  heating of a rarefied gas in a channel}},\ }\href@noop {} {\bibfield
  {journal} {\bibinfo  {journal} {Journal of Fluid Mechanics}\ }\textbf
  {\bibinfo {volume} {800}},\ \bibinfo {pages} {433} (\bibinfo {year}
  {2016})}\BibitemShut {NoStop}%
\end{thebibliography}
%

\end{document}